\documentclass{emulateapj}

\usepackage{rotating}

\usepackage{color}
\definecolor{excol}{rgb}{0.1,0.6,0.1}
\definecolor{rcol}{rgb}{0.2,0.2,0.8}

\newcommand{\msun}{M$_\odot$}
\newcommand{\HI}{\ion{H}{1$~$}}
\newcommand{\sbar}{\ensuremath{\bar{S}}}

\shorttitle{NGC~2903 and its Environment}
\shortauthors{Irwin et al.}

\begin{document}


\title{$\Lambda$CDM Satellites and \HI\ Companions ---\\
The Arecibo ALFA Survey of NGC~2903}


\author{Judith A. Irwin}
\affil{Department of Physics, 
Engineering Physics and Astronomy, Queen's University,
    Kingston, ON, Canada, K7L 3N6}
\email{irwin@astro.queensu.ca}
\author{G. Lyle Hoffman}
\affil{Department of Physics, Lafayette College, Easton, PA, 18042}
\email{hoffmang@lafayette.edu}
\author{Kristine Spekkens}
\affil{Department of Physics, Royal Military College of Canada, PO Box 17000,
    Station Forces, Kingston, ON, Canada K7K 7B4}
\email{kristine.spekkens@rmc.ca}
\author{Martha P. Haynes\altaffilmark{1}}
\affil{Center for Radiophysics and Space Research, Cornell University,
    Ithaca, NY, 14853, USA}
\email{haynes@astro.cornell.edu}
\author{Riccardo Giovanelli\altaffilmark{1}}
\affil{Center for Radiophysics and Space Research, Cornell University,
    Ithaca, NY, 14853, USA}\email{riccardo@astro.cornell.edu}
\author{Suzanne M. Linder}
\affil{Hamburger Sternwarte, Universit\"at Hamburg,
Gojenbergsweg 112, D-21029, Hamburg, Germany}
\email{suzanne.linder@hs.uni-hamburg.de}
\author{Barbara Catinella\altaffilmark{2}}
\affil{Max-Planck-Institut f{\"u}r Astrophysik, D-85748 Garching,
Germany}
\email{bcatinel@MPA-Garching.mpg.de}
\author{Emmanuel Momjian\altaffilmark{3}}
\affil{NAIC-Arecibo Observatory, HC3 Box 53995, Arecibo, PR, 00612, USA}
\email{emomjian@naic.edu}
\author{B{\"a}rbel S. Koribalski}
\affil{Australia Telescope National Facility, CSIRO, Epping, NSW 1710, 
Australia}
\email{Baerbel.Koribalski@csiro.au}
\author{Jonathan Davies}
\affil{School of Physics and Astronomy, Cardiff University of Wales, Cardiff,
CF24 3YB, UK}
\email{jonathan.davies@astro.cf.ac.uk}
\author{Elias Brinks}
\affil{Centre for Astrophysics Research, Science and Technology Research
Institute, University of Hertfordshire, Hatfield, AL10 9AB, UK}
\email{ebrinks@star.herts.ac.uk}
\author{W. J. G. de Blok}
\affil{Department of Astronomy, University of Cape Town, Rondebosch 7700, South
Africa}
\email{edeblok@ast.yct.ac.za}
\author{Mary E. Putman}
\affil{Department of Astronomy, Columbia University, New York, NY, 10027, USA}
\email{mputman@astro.columbia.edu}
\and
\author{Wim van Driel}
\affil{Observatoire de Meudon, 5 Place Jules Janssen, 92195 Meudon, France}
\email{wim.vandriel@obspm.fr}
\altaffiltext{1}{National Astronomy and Ionosphere Center,
Cornell University, Ithaca, NY, 14853.}
\altaffiltext{2}{NAIC-Arecibo Observatory, HC3 Box 53995, Arecibo, PR, 00612, USA}
\altaffiltext{3}{NRAO, PO Box O, Socorro, NM, 87801, USA}

\begin{abstract}

We have conducted a deep, complete \HI survey,
using Arecibo/ALFA, of a field 
centered on the nearby, isolated galaxy, NGC~2903,
which is similar to the Milky Way in its properties.
The field size was 150 kpc $\times$ 260 kpc and the 
final velocity range spanned
from 100 to 1133 km s$^{-1}$.
The ALFA beams have been mapped as a function of azimuth
and cleaned from each azimuth-specific cube prior to forming final
cubes.
 The final \HI data
 are sensitive down to an \HI mass of
2$\,\times\,10^5$ M$_\odot$ and column density of
$2\,\times\,10^{17}$ cm$^{-2}$ at the 3$\,\sigma\,$2$\,\delta\, V$ level,
where $\sigma$ is the rms noise level and $\delta\,V$ is the
velocity resolution.
 NGC~2903 is found to have an \HI 
envelope that is larger than previously known, extending to at least 3 times
the optical diameter of the galaxy.  Our search for companions 
yields one new discovery with
an \HI mass of $2.6\,\times\,10^6$ M$_\odot$.  The companion is
64 kpc from NGC~2903 in projection, 
is likely associated with a small optical galaxy of similar
total stellar mass, and is dark matter dominated, with a 
total mass $>\,10^8$ M$_\odot$.  In the region
surveyed, there are now two known companions: our new discovery and a previously
known system that is likely a dwarf spheroidal, lacking \HI content.
If \HI constitutes 1\% of the total mass in all possible companions, then
we should have detected 230 companions, according to 
$\Lambda$CDM predictions.  Consequently, if this number of
dark matter clumps are indeed
 present, then they contain less than 1\% \HI content, possibly existing as
very faint dwarf spheroidals or as starless, gasless dark matter clumps.

\end{abstract}


\keywords{galaxies: individual (NGC~2903) --- galaxies: spiral --- galaxies:
formation --- cosmology: dark matter --- radio lines: ISM }



\section{Introduction}
\label{introduction}

Weak gravitational lensing studies have shown that stellar light traces dark
matter on supercluster and cluster scales \citep{hey08}. 
The issue is much less clear
 on sub-galactic scales, however, as evidenced by
the well-known `missing satellites' problem around the Milky Way (MW).
At issue is the fact that
 the predicted number of satellites
based on Cold Dark Matter (CDM) and $\Lambda$ Cold Dark Matter 
($\Lambda$CDM) simulations of galaxy formation is significantly greater
than the observed number of dwarf MW companions 
\citep{kau93, kly99, moo99,
die07b}.

A number of explanations for this discrepancy have been proposed.  These
include the suppression of star formation due to the
ionization of the gas \citep{bar99, ben02, sha03, gne08}
or dissociation of molecular hydrogen \citep{hai97},
gas-stripping due to supernovae-driven winds from an early star-formation
epoch
 \citep{hir98, kly99},
only the most massive halo substructures forming stars
\citep{sto02},
 the disruption of satellites by tidal
stripping/stirring 
\citep{kly99, may01a, may01b, kra04},
 confusion between dwarf satellites and high velocity
clouds (HVCs, see Klypin et al. 1999 \nocite{kly99} for a summary),
the suppression of small-scale power in simulations 
via contributions from Warm Dark Matter (Avila-Reese et al.
2001, Zentner \& Bullock 2002)\nocite{avi01, zen02},
and incompleteness in the census of MW satellites \citep{mat98}.

It is now clear that the latter explanation has played a part, since
in the past few years, the known population of dwarf MW satellites has
almost doubled based on scrutiny of the Sloan Digital Sky Survey
(SDSS, York et al. 2000\nocite{yor00}, see also
Koposov et al. 2008\nocite{kop08}); however, although the detection of the
new satellites alleviates the problem, it does not eliminate it entirely
since, after making corrections for sky coverage, the discrepancy is still
approximately a factor of 4
\cite[see][]{sim07}.

\begin{figure*}
\plotone{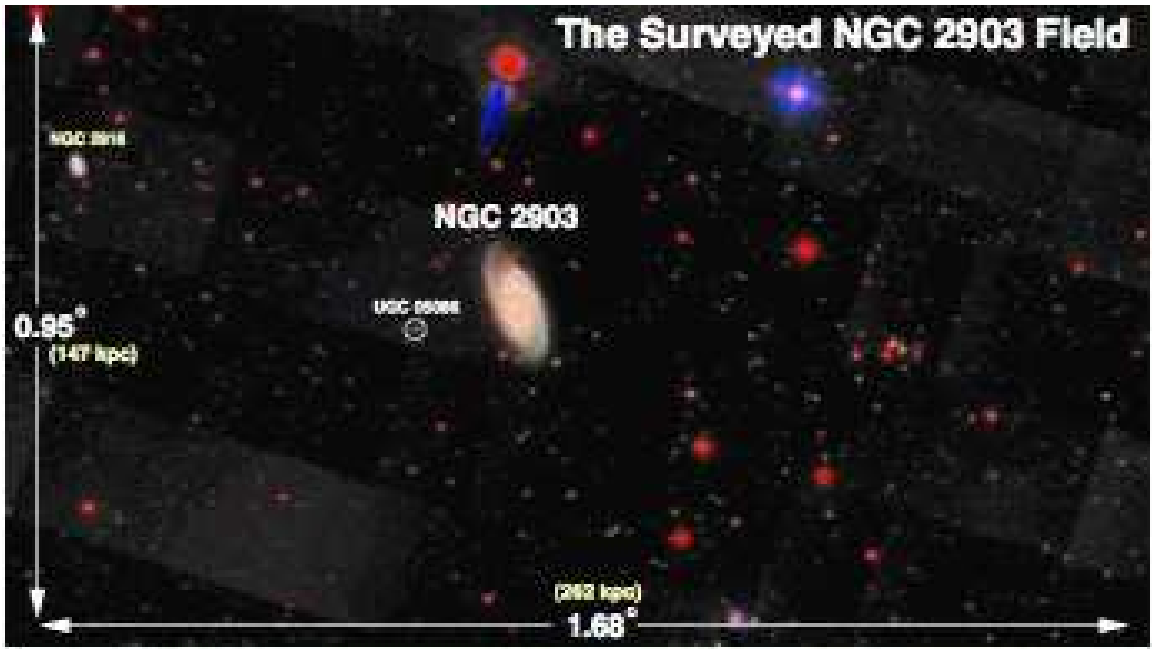}
\caption{Sloan Digital Sky Survey (SDSS DR6) image of the field around
NGC~2903 that has been surveyed by the Arecibo telescope.
 NGC~2903 and its companion, UGC~5086,
are labelled, as well as the scale.  The background galaxy,
NGC~2916, is also labelled
(see Sect.~\ref{companions}).
\label{ngc2903_colour}}
\end{figure*}

The concept that star formation might, in some way, have been suppressed in 
systems of low total mass, suggests the possibility that
dark matter substructure could still be traced by atomic hydrogen
(\ion{H}{1}) 
even though appreciable stellar content may be missing.  However, 
many searches, of variable sensitive and coverage, have
been undertaken for low mass starless companions, with
little success (see Sect.~\ref{previous}) and some have suggested
that small galaxies which retain \HI are likely to have developed stars
as well
\citep{bri04, tay05}.

The advent of the 7-beam Arecibo L-band Feed Array (ALFA, see
Giovanelli et al. 2005a\footnote{See also 
{\tt http://www.naic.edu/alfa/}}\nocite{gio05})
has now provided an opportunity to survey nearby
systems for the possible presence of such `dark companions', with
unprecedented sensitivity, coverage, and speed.
We report here the results of a targeted deep survey of a
 single, isolated galaxy, \object{NGC~2903} (see 
Sect.~\ref{ngc2903}).
In this paper, we outline 
our observational procedure and data reduction, we present 
global results for NGC~2903 and then concentrate on companions and
their implications for primordial dark matter searches.  
  The details of the \HI in
NGC~2903, itself, will be left for a subsequent paper.  
Please note that data related to this project, including some software
that we have developed,
data from related observations, our final cubes and beam maps
 can be found on our NGC~2903
website\footnote{\tt http://www.astro.queensu.ca/$\,\tilde{\,\,}$irwin/ngc2903\label{footnote2}}.

In
Sect.~\ref{previous}, we outline previous surveys that have taken
place, presenting a comparison with our approach and,
in Sect.~\ref{ngc2903}, we discuss NGC~2903
and its environment.  Since this paper introduces new techniques
for observing and reducing Arecibo/ALFA data, we discuss these in
detail in Sects.~\ref{observations} and
Sect.~\ref{data_reduction}.  
Our detection thresholds and data quality are given in
Sect.~\ref{data_quality}, 
Sect.~\ref{results}
presents the results for NGC~2903 and its environment, and
Sect.~\ref{discussion} and Sect.~\ref{conclusions} provide the
discussion and conclusions, respectively.

\section{Previous \HI Surveys and Comparison with NGC~2903}
\label{previous}

Current observational data suggest that \HI clouds tend not to
be `intergalactic', but rather associated with galaxies
\citep{bri04}.
Various searches for faint \HI around galaxies, however,
 have typically been hampered by 
the need to choose sensitivity at the expense of coverage or vice versa.

For wide coverage, the blind \HI Parkes All Sky Survey
\citep[HIPASS][]{bar01, kor04, mey04} 
revealed only one definite extragalactic \HI cloud 
in the NGC~2442 group with a high \HI mass of about 10$^9$ \msun\ 
\citep{ryd01}.  This cloud has been interpreted as a remnant of
a tidal interaction \citep{bek05} and cannot be considered primordial.

As for targeted searches, lower mass limits have been achieved. 
Zwaan (2001)\nocite{zwa01} and de Blok (2002)\nocite{deb02} 
made incomplete samplings of several galaxy groups to limits of a
few $\times\,10^6$ \msun.
Minchin et al. (2003)\nocite{min03}
surveyed  
the Cen A group to a 
limit of
$2\,\times\,10^6$ \msun, 
and Pisano et al. (2004, 2007) \nocite{pis04}\nocite{pis07}
observed 6 Local Group  analogs
to 
$2\,-\,5\,\times\,10^6$ \msun.
Kova{\u{c}} et al. (2005)\nocite{kov05} 
 completely surveyed the Canes Venatici group
to a limit of $10^6$ - $10^7$ \msun.
Barnes \& de Blok (2004)\nocite{bar04} searched for faint \HI companions around
NGC~1313 and Sextans A to $\sim$ 10$^6$ M$_\odot$.
Pisano \& Wilcots (1999, 2003) \nocite{pis99} \nocite{pis03}
searched for gas rich companions around 6 isolated galaxies to an
approximate detection limit of only $10^8$ \msun\footnote{Note that the various
groups have not all used the same criteria for determining their mass limits.}.
In these and other targeted surveys \citep[see also][]{kil06}, 
neither starless \HI companions nor HVCs, where the sensitivity was
sufficient \citep[e.g.][]{pis07},
 were detected with the exception of \HI that could again be 
attributed to tidal debris \citep[e.g. see][]{bek05}.  


In contrast to the targeted searches
described above, our study of a relatively isolated system
(see Sect.~\ref{ngc2903}) simplifies the 
interpretation of any \HI detections since tidal explanations are
much less likely.  Moreover,
use of the 305 m diameter Arecibo radio telescope has placed
these observations among the most sensitive yet achieved
(see Sect.~\ref{noise}). 
While slightly lower \HI mass limits have been claimed 
in deep interferometric observations of 
NGC~891 (Oosterloo et al. 2007\nocite{oos07}),
M~31 and M~33 (Westmeier et al. 2005\nocite{wes05a}), our
combination of low mass detection limits, the lowest \HI column
density limits
yet achieved in such studies, the sensitivity to broad scale structure
not possible via interferometers, and the complete (and large) sky
coverage combine to make this survey unique.


\section{NGC~2903 and its Environment}
\label{ngc2903}

NGC~2903 (Table~\ref{table1} and Fig.~\ref{ngc2903_colour})
has a number of assets that make it a good target for
deep \HI mapping. It falls within the
declination range of the Arecibo telescope, it is bright and massive
so there is a reasonable expectation of the presence of 
$\Lambda$CDM (or other) satellites, it is of large angular size
so is easily resolved by the Arecibo beam, it is
nearby yet {lies beyond the Local Group} 
({$D$} = 8.9 Mpc; Drozdovsky \& Karachentsev 2000\nocite{dro00};
$1^\prime\,=\,2.6$ kpc \footnote{Literature values range from
6.01 Mpc to 11.65 Mpc, depending on corrections for
local motions (see the NASA/IPAC Extragalactic Database (NED)).}),
and some previous \HI observations of
the galaxy are available for comparison 
\cite{beg87, beg91, hew83, hay98}.
An important characteristic is that it is non-interacting and 
isolated, in the sense that no galaxies larger than one quarter
of its optical size are present within 20 optical diameters away
(No. 0347 in the
Catalogue of Isolated Galaxies, Karachentseva 1973\nocite{kar73}
\footnote{At the time of writing, NGC~2903 has not
been included in the Analysis of the interstellar Medium
of Isolated Galaxies (AMIGA) catalogue \citep[e.g.][]{ver07} due to
its large angular size.}, see also Haynes et al. 
 1998\nocite{hay98}).

NGC~2903 is characterized by its barred, grand-design spiral
pattern.  It displays a number of `hot spots'  
in its nuclear region as well as a ring of star formation 
 \citep[e.g.][]{per00}.  The nuclear dust distribution is
chaotic \cite{mar03}.  The CO emission is concentrated along
the bar \citep{reg99} and the star formation rate (SFR)
per unit area 
is enhanced by an order of magnitude in the nucleus in comparison
to the disk \cite{jac91}. 
A soft X-ray halo extending to the west of the nucleus has been
interpreted as outflow from a nuclear starburst-driven wind 
\cite{tsc03}.
  
Aside from evidence of nuclear star formation, however, 
 NGC~2903 is a typical massive spiral
whose properties are similar to those of the Milky Way.  Its
global SFR (2.2 \msun yr$^{-1}$, Table~\ref{table1}) 
is comparable to the MW value
($\sim$ 4 \msun yr$^{-1}$,
Diehl et al. 2006\nocite{die06}), considering the different
methods for estimating this value.  More importantly, its rotation
curve  shows a rise to 210 km s$^{-1}$ at
a galactocentric radius of R$\,\sim\,$4 kpc, declining slightly to 
180 km s$^{-1}$ by R$\,\sim\,$33 kpc, its outermost measured point
\citep{beg91}.  These values
agree with the rotation curve of the Milky Way over
4$\,\le\,$R (kpc)$\,\le\,$33 to within error bars \citep{xue08}.
Aside from environment, therefore, NGC~2903 appears to be an analog
of the MW.

As indicated above, NGC~2903 is considered to be an isolated
galaxy, given the dearth of nearby companions sufficiently massive
to perturb it. 
 However, two small
companions, UGC~5086 and D565-06, are known to be associated
\citep{dro00} and, from an optical search for additional possible
companions within similar radii, we have now identified a third 
companion,
D565-10.  D565-10 was found from a search
over the spatial region and velocity range within which UGC~5086
and D565-06 have previously been found.  Since its 
separation from NGC~2903 in both position and velocity space is
less than that of D565-06, we include it as a newly identified
companion here.
These three galaxies and their known properties
are listed in Table~\ref{table2}.  Of the three, only UGC~5086 is within
our surveyed 
field of view and is labelled in
Fig.~\ref{ngc2903_colour}.  

Aside from the \HI observations listed above, more recent \HI data
 from the HIPASS  
\cite{won06} and the Westerbork SINGS survey
\cite{bra07} are now also available. 
NGC~2903 is also in The \HI Nearby Galaxy
Survey (THINGS, Walter et al. 2008\nocite{wal08}) which makes use of
Very Large Array (VLA) data.
 At the time of our observations, five archival
VLA unpublished \HI data sets were available, 
all of which we have 
reduced. Of these, two sets produced good data.  
These are: a) observing run AO125, taken 29 Sept. 1996 constituting
2.03 hours on source in D configuration, and b) run 
AW536, taken 22 Apr. 2000, constituting 2.68 hours on source
 in C configuration.  
We do not reproduce the VLA cubes here, but make them available
on our NGC~2903 website (see Footnote~\ref{footnote2}),
 and refer to them, as needed, only for
comparison purposes.  The VLA data sets are of higher spatial
resolution than the Arecibo/ALFA data, but are much less
sensitive (see Sect.~\ref{envelope}, for
example).   These reference VLA data sets  
predate those of THINGS\footnote{The THINGS data
set achieves a sensitivity of
N$_{HI}\,=\,4\,\times\,10^{19}$ cm$^{-2}$ at a resolution
of 30 arcsec, using the same criteria as we will set out in
Sect.~\ref{data_quality}.\label{things_footnote}}.


\begin{table}[h]
\begin{center}
\caption{Parameters of NGC~2903\tablenotemark{a}\label{table1}}
\begin{tabular}{lc}
\tableline\tableline
Parameter & NGC~2903 \\
\tableline
R.A. (h m s) & $\,\,~$09 32 10.11   \\
Decl. ($^\circ$ $^\prime$ $^{\prime\prime}$) & $\,$21 30 03.0  \\
Type & SB(s)d \\
V$_\odot$ (km s$^{-1}$)\tablenotemark{b} & 556.0  \\
D (Mpc)\tablenotemark{c}  & 8.9   \\
2a $\times$ 2b ($^\prime~\times~^\prime$)\tablenotemark{d}& 12.6
$\times$ 6.0 \\
$\,$~~~~~(kpc $\times$ kpc) & 32.6 $\times$ 15.5 \\
L$_{FIR}$ (10$^{10}$ L$_\odot$)\tablenotemark{e} & 1.30 \\
L$_{IR}$ (10$^{10}$ L$_\odot$)\tablenotemark{e} & 1.80 \\
SFR (\msun yr$^{-1}$)\tablenotemark{f}& 
2.2 \\
\tableline
\end{tabular}
\tablenotetext{a}{Data from NED
unless otherwise indicated.}
\tablenotetext{b}{Heliocentric velocity.}
\tablenotetext{c}{Drozdovsky \& Karachentsev (2000)\nocite{dro00}.}
\tablenotetext{d}{Optical major $\times$ minor axis diameters. The semi-major\\
diameter, $a$, is equivalent to $R_{25}$, the radius in the \\
$B$-band
 at the 25.0\,mag\,arcsec$^{-2}$ isophote level.}
\tablenotetext{e}{Infra-red and far infra-red luminosity 
\citep{san03}, \\
adjusted to our distance.}
\tablenotetext{f}{Star formation rate from L$_{FIR}$ 
and the formalism of \\
Kennicutt (1998)\nocite{ken98}
.}
\end{center}
\end{table}

\begin{table}[h]
\begin{center}
\caption{Companion Galaxies of NGC~2903\tablenotemark{a}\label{table2}}
\begin{tabular}{lccc}
\tableline\tableline
Parameter & UGC~5086$^b$ & D565-06$^c$ & 
D565-10$^c$\\
\tableline
R.A. (h m s) & 9 32 48.9   & 9 19 30.0 & 9 30 12.8 \\
Decl. ($^\circ$ $^\prime$ $^{\prime\prime}$) & 21 27 55 & 21 36 12 & 19 59 26 \\
Sep. ($^\prime$)\tablenotemark{d}& 9.3 & 176.8 & 94.7\\
~~~~~(kpc)\tablenotemark{d} & 24 & 458 & 245 \\
V$_\odot$ (km s$^{-1}$)\tablenotemark{e} & 510 $\pm$ 30$^f$ & 498 $\pm$ 2 &
562 $\pm$ 1\\
2a $\times$ 2b ($^\prime\times^\prime$)\tablenotemark{g}
& 0.9 $\times$ 0.9 & 0.7 $\times$ 0.6 & 0.7 $\times$ 0.6\\
Mag.\tablenotemark{h} & 18  & 16.95 & 17.0 \\
Ref.\tablenotemark{i} & DK00 & DK00 & this work \\
\tableline
\end{tabular}
\tablenotetext{a}{Data from NED unless otherwise indicated.}
\tablenotetext{b}{Alternate names: D565-05 in the Low Surface Brightness\\
 Galaxy
Catalogue (LSBC); J093248.81+212756.2 in the SDSS.}
\tablenotetext{c}{Identifier in the LSBC.}
\tablenotetext{d}{{Projected s}eparation from the center of NGC~2903.}
\tablenotetext{e}{Heliocentric radial velocity.}
\tablenotetext{f}{Optical velocity, $c z$, as given in the SDSS Data
Release 6 (DR6).}
\tablenotetext{g}{Major axis $\times$ minor axis diameters.}
\tablenotetext{h}{Optical B magnitude.}
\tablenotetext{i}{Reference for association with NGC~2903.  \\ DK00 = 
Drozdovsky \& Karachentsev (2000)\nocite{dro00}}
\end{center}
\end{table}

\section{Observations}
\label{observations}

Observations were carried out with the 305 m telescope of the Arecibo
Observatory\footnote{The Arecibo Observatory is part of the 
National Astronomy and Ionosphere Center (NAIC), a national 
research center operated by Cornell University under a cooperative 
agreement with the National Science Foundation (NSF).}
using the 7 beam 
ALFA receiver system 
\citep[see Fig.~2 of][for the ALFA
beam geometry]{gio05} with
 the Wideband Arecibo Pulsar Processor
(WAPP) back-end spectrometer system. 
The total observing time allocated for this project was 97 hours,
divided into 37 observing blocks (November 28-30, December 1-6, 14-23,
26 2004; February 10-13, 28, March 1-6, 21-26 2005) 
{carried}
out during the commissioning phase of ALFA. 
{ The observing setup is summarized in Table~\ref{table3}.}

{Because the ALFA beams can have coma lobes as high as 20\% (7 dB),
high-sensitivity observations of extended objects with this instrument must
account for contributions from stray/unwanted radiation into these lobes;
that is, we obtain a ``dirty map'' which must be ``cleaned''.  
Our basic approach is
therefore to map the field in Fig.~\ref{ngc2903_colour} as well as the 7 ALFA
beams in a fixed number of telescope configurations. The beam maps are used
to deconvolve the sidelobe contribution to the galaxy map in each
configuration after which
 these clean maps are combined to form our final datasets. In
Sects.~\ref{galaxy_mapping}~and~\ref{beam_mapping}, we describe our observing
strategy for mapping the galaxy and the beams, respectively.}

\subsection{Observations of NGC~2903}
\label{galaxy_mapping}

The mapping of NGC 2903 was conducted in a ``Fixed Azimuth Drift'' mode { similar to that} adopted in the
Arecibo Galaxy Environment Survey
\citep[AGES,][]{aul06}\footnote{\tt http://www.naic.edu/$\,\tilde{\,\,}$ages}.
It is, however,
somewhat less efficient than the {dual-pass} strategy 
employed by the Arecibo Legacy Fast ALFA Survey
(ALFALFA, Giovanelli et al. 
2005a\nocite{gio05})\footnote{\tt http://egg.astro.cornell.edu/alfalfa/}, 
{mainly because we observed NGC~2903 at 12 separate azimuths in order 
to increase the effective integration time per point to 6 times the 
ALFALFA value.} 


For each of 12 azimuth{s ($AZ = 104^\circ$,
 107$^\circ$,  109$^\circ$, 116$^\circ$,  128$^\circ$,
 153$^\circ$, 200$^\circ$,  230$^\circ$,
 245$^\circ$,  250$^\circ$, 253$^\circ$, 255$^\circ$),
ALFA}  
was positioned to point $4^m$ in R.A.
ahead of the source, then held motionless while the source drifted through.
Spectra for each linear polarization of each of the {7} feeds were 
recorded at a rate of once per second during the drift scan.  
Thus 12 drift scans centered at a single declination could optimally be 
obtained in one observing session.  The azimuths 
were determined by calculating the minimum 
{time} required for slewing and
resetting/adjusting system parameters in preparation for the next drift
at the next azimuth, thus minimizing overheads.

Before each drift scan at each azimuth, 
the ALFA {turret} was rotated so as to produce 
equal separations in declination  between successive beams {(but see below)}.
The azimuth and zenith angle were also both adjusted so that 
{a} drift would cross the same declination (J2000) 
any time {it} was repeated.  

{On a given night, drifts at all 12 azimuths} 
kept the center
beam of ALFA at the same declination. On following nights, 
the declination was shifted by $4\arcmin45\arcsec$ so that
 the center beam drifted through the interstice between the northermost beam 
of the preceding night's drift and the southernmost of the following.
This helped to produce a map with as uniform a sensitivity as possible 
{and Nyquist-sampled the survey area in declination.}

In the allocated time we were able to obtain full sets of 12 drifts 
each, across 
13 nearly constant declination tracks with the center beam spanning 
$21\arcdeg03\arcmin56\arcsec$ to $22\arcdeg00\arcmin56\arcsec$ (J2000),  
and with another 
full set of 12 drifts spanning $21\arcdeg13\arcmin26\arcsec$ to 
$21\arcdeg27\arcmin41\arcsec$ and $21\arcdeg37\arcmin11\arcsec$ to
$21\arcdeg51\arcmin26\arcsec$. (The number of drifts at a given declination,
referred to as $N_p$, is given in Fig.~\ref{noise_fig}d)
Compared to these repeated drift ranges, our final map is undersensitive 
in the central declination strip 
$21\arcdeg32\arcmin26\arcsec$, the two southernmost strips and the two
 northernmost strips (see Sect.~\ref{noise} for sensitivity).

The {F}ixed  {A}zimuth {D}rift mode
introduces some variations in declination. First of all, since the telescope was held at fixed 
azimuth and  zenith angle through each drift, the declination 
tracked by each beam changed by a small amount from the start 
to the end of the drift. 
{We have computed the average variation in beam position from this effect,} and find it to be less than 1 arcsec and therefore negligible.
Secondly,
the declination spacing between ALFA beams {at a given azimuth} 
was not exactly uniform.  
The variation in {beam} spacing is
typically { $\approx2\% $} of the beam size, a value comparable to
the $\approx\,$5$^{\prime\prime}$ pointing accuracy of the Arecibo telescope.
Finally, the declination spacings vary with azimuth {due to the elliptical illumination pattern of the Arecibo telescope}.
These declination separations varied {monotonically} from 
{$\sim123\arcsec$  $AZ=180\arcdeg$ 
to $\sim108\arcsec$ at $AZ=104\arcdeg$ and $AZ=255\arcdeg$. 
We account for the latter two effects in the data reduction 
(see Sect.~\ref{galaxy_calibration}).}


\begin{table}[ht]
\begin{center}
\caption{Observing Setup\label{table3}}
\begin{tabular}{lc}
\tableline\tableline
Parameter & Value \\
\tableline
{\bf Galaxy} & \\
\tableline
Bandwidth (MHz) & 12.5  \\
Center frequency, $\nu_c$ (GHz) & 1.417 \\
No. Channels & 2048 \\
Velocity Coverage 
(km s$^{-1}$)\tablenotemark{a}
 & $-$585.5 $\to$ 2064.2 \\
Channel width (km s$^{-1}$)& 1.293 \\
Drift Scan Duration (s) & 480 \\
\tableline
{\bf Beams}&\\
\tableline
Bandwidth (MHz) & 50.0  \\
Center frequency (GHz) & 1.405 \\
Drift Scan Duration (s) & 240 \\
\tableline
\end{tabular}
\tablenotetext{a}{All velocities in this paper are heliocentric.}
\end{center}
\end{table}


\subsection{Beam Mapping}
\label{beam_mapping}

\begin{sidewaysfigure*}
\epsscale{1.0}
\plotone{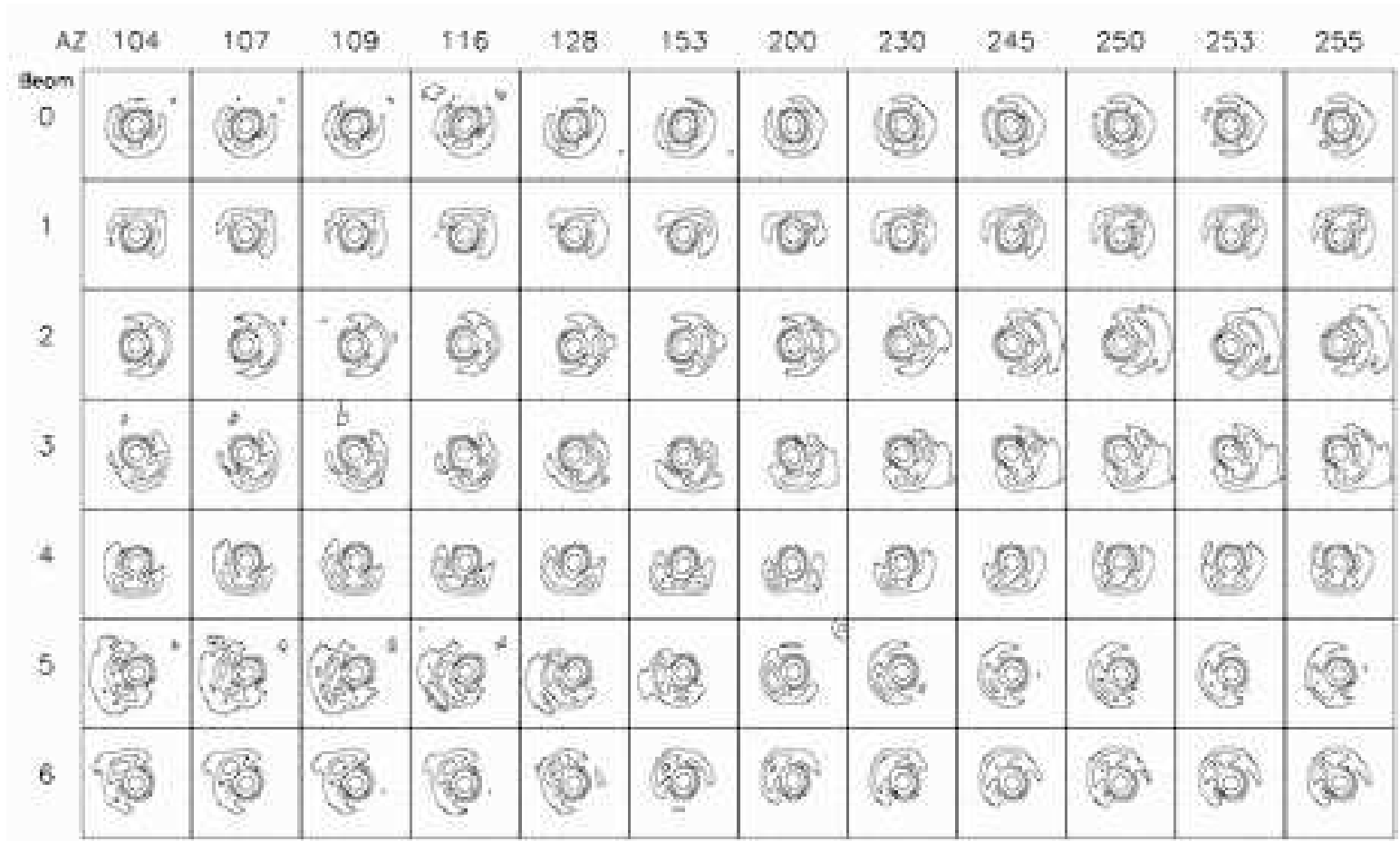}
\caption{Maps of each ALFA beam (rows) used to {deconvolve} sidelobe and 
stray radiation contributions from NGC~2903 observations at each azimuth 
(columns). Each panel spans 24\arcmin\ in both R.A. and decl., and contours
 are at (-18, -12, -9, -3) dB. {See fig.~2 of Giovanelli et al.\ 2005a 
 for an illustration of the beam locations in the ALFA footprint.} FITS files containing the 84 beams {may} be downloaded from {our} website. 
\label{bw}
}
\end{sidewaysfigure*}

The outer beams of ALFA have significant coma lobes, and the contributions of 
stray radiation and outer sidelobes to all the beams is not negligible. While precise {beam} maps were produced 
by Cort{\'e}s (2003)\nocite{cor03}, they do not {account for variations with azimuth {and} zenith angle or blockage by the platform and cables.} 

Consequently it was necessary to map the beams by observing strong 
unresolved
continuum sources 
using a mapping strategy similar to that {described in Sect.~\ref{galaxy_mapping}.} 
In the case of the beams, however, we chose twice as
many azimuth settings (24 instead of 12) and shorter drift scans
 (Table~\ref{table3}).

The declination of Beam 0, which is the center beam of ALFA,
was shifted by $1\arcmin53\arcsec$
 (about half a beamwidth) from
one night to the next, and we completed 13 separate drifts on 
each source.
No single source of sufficient strength could be mapped at each 
of the 24 azimuths in the time we were allotted on any one night, 
so we mapped two {of them}:  
J080538+210651 (0.9 Jy) at  $AZ \lesssim 120\arcdeg$ and 
J102155+215931 (1.7 Jy) at $AZ \gtrsim 120\arcdeg$.

This procedure gave us beam maps approximately $56\arcmin$ long 
in R.A.  and spanning  $23\arcmin$
in decl.
Only Beam 0 is mapped to equal distances north and 
south of the center, however.
While the span in R.A. is sufficient to map the first several 
sidelobes of each of the outer beams, the span in declination 
falls short of reaching the second sidelobe for the northernmost 
and southernmost beams.  This will be discussed further in
Sect.~\ref{beam_reduction}.

\section{Data Reduction and Processing}
\label{data_reduction}

The drift scans were bandpass-subtracted and baseline-flattened using  
Interactive Data Language (IDL) 
procedures 
written by P. Perillat for general use at Arecibo Observatory, 
and by R. Giovanelli and B. {R.} Kent for the ALFALFA precursor 
data \citep{ALFALFA2}.
Preliminary calibration was accomplished using the system's equivalent
flux density 
for each beam as a function of zenith angle, provided by 
Arecibo Observatory staff.  Data reduction specific to the galaxy and
beams is described {below}.

\subsection{{Galaxy Data Reduction}}
\label{galaxy_calibration}

Initial maps of
NGC~2903 from the drifts 
showed significant striping across the galaxy, indicating that the 
calibration of the individual beams was not sufficiently precise.
An attempt to improve the calibration by integrating over the Galactic 
\ion{H}{1} emission, requiring the integral to be the same for each beam, 
proved unsuccessful since the separate beams follow different tracks 
across the Galactic emission and, on the scale of the NGC~2903 map, 
there is significant variation of the Galactic \ion{H}{1} 
emission between those tracks.

To improve the calibration, we sought to make use of the continuum 
sources in the mapped field.
{Sources} with sizes small compared
 to the Arecibo beam and with peak flux densities exceeding 10 mJy
 {were selected from NED}.
Gaussians were fitted to {corresponding} detected continuum signals {in our data}, 
the distance from the {fitted} peak to the 
catalogued source position was determined, and the ratio 
of the flux expected at that position to that observed was calculated.
Only detections that fell within half a beamwidth of the source were used.
Starting with the strongest source, then working down the list in order 
of source strength, the factors by which each beam must be multiplied 
to place their gains on a common scale were determined.  
The resulting calibration is referenced to the flux of the strongest 
source in the field, J093215+211243, which we took to have peak 
flux density 562.3 mJy at the time of our observations.  
This approach significantly reduced the striping {in the final maps}, 
though we note 
{the presence of residuals 
that will be discussed in Sect.~{\ref{residual}}.}

The drifts for each azimuth separately were then gridded into 
a datacube with axes, right ascension (R.A.), declination (Decl.), 
and {velocity (V)}.
This was done using the ALFALFA IDL gridding tool 
(Giovanelli et al. 2008 in prep.)
modified for the NGC~2903 drift length and calibration method.
At each defined point in the grid, the gridding tool produced a 
weighted average of emission from nearby one-second spectral samples, 
using the positions recorded in the data headers for each.
No assumption of constant separation in declination from one beam 
to the next was required in this process.
The beam which dominated the weight for each grid point was also 
recorded in the data structure, for use later by the cleaning software
(see Sect.~\ref{cleaning}).
The grid pixel size was 30$^{\prime\prime}$ in R.A. and Decl., and the velocity range
was chosen to extend well beyond 
{that of} NGC~2903, {excluding Galactic emission.}

There was no significant radio frequency interference from outside the 
observatory in our spectra.  However, there was an internally generated
``wandering birdy'' \citep{ALFALFA2} present in some of our drifts.
This was an interference spike that drifted non-monotonically 
in frequency during particular drifts.
The source has since been identified and corrected, 
and has not been seen in any observations since early 2005, to our knowledge.
Fortunately the birdy's wanderings did not take it close to the 
NGC~2903 velocity range during our observations for the most part.
In those few spectra where it would have caused difficulty 
in the cleaning process, it was excised by interpolating 
between spectral channels just outside the spike. 

\subsection{{Beam Data Reduction}}
\label{beam_reduction}

To produce a separate 2-dimensional (2D) map of each beam at each 
azimuth, we extracted 
the individual beam drifts from the sets of drift scans, 
then constructed maps of the continuum source as seen by each beam, 
producing 7 maps for each of the 24 azimuth settings.
The continuum maps were written out from IDL into
Flexible Image Transport System
(FITS) format, then read into the 
Astronomical Information Processing System software, 
{aips}++ \footnote{See http://aips2.nrao.edu},
of the National Radio Astronomy Observatory (NRAO).
Standard 2D gaussian fitting routines in {aips}++ were used to fit 
and subtract out each catalogued continuum point source outside 
the much stronger target source.
Noticeable sidelobe signal from these non-target sources was 
zeroed as well, as long as there was no confusion with the 
sidelobes of the target source.
Our inability to completely erase these extraneous sources 
limits {the dynamic range of the final 
galaxy maps 
(see Sect.~\ref{residual}).}



These maps were then read into the Astronomical Image Processing 
System (AIPS)\footnote{http://www.aips.nrao.edu/} 
for further processing.  The goal was to create
{a single} map for each of the 7 beams at {each of the}
12 galaxy azimuths (84 beam maps in total).  To do this,
the beam maps were placed on the same 
amplitude scale and beam maps at azimuths {adjacent to} 
the galaxy azimuth were averaged together, weighted {by distance} from the
 {galaxy} azimuth.  
For example, beam {maps}
at {$AZ=103^\circ$ and $AZ=106^\circ$} were {averaged to obtain an estimate of the beam at the galaxy azimuth $AZ=104^\circ$},
with a higher weight {attributed to $AZ=103^\circ$}.    

The final beam maps {at each galaxy azimuth} covered 54$^\prime$ in {R.A.} and 21$^\prime$ in {decl}.  This was more than
sufficient to fully map the beam shape in {R.A.}.  However,
as indicated in Sect.~\ref{beam_mapping}, the {decl.} span
does not reach the second sidelobe for the northernmost 
and southernmost beams. In addition, 
 for 4 of the 7 beams,
the first sidelobe on one side only was cut off in {decl.}
at approximately the midpoint of its peak.  For these cases,
to avoid the introduction of artifacts during the {deconvolution} process
(Sect.~\ref{cleaning}),
the beam sidelobe was extended/smoothed at the edge by a gaussian of width, 
2.3$^\prime$.  

All 84 beams are displayed in Fig.~\ref{bw} and nicely show the changing
beam structure with changing a{z}imuth\footnote{FITS files for these beams
are available {on} the NGC~2903 website.}.  These beams were then used for
the IDL-based clean described in the next section.

\subsection{{Image Deconvolution and Final Cubes}}
\label{cleaning}

To achieve high sensitivity to low-level emission from the outer edges of the
 galaxy it {is} necessary to remove {the}
sidelobe and stray radiation
 contribution to the maps of NGC~2903. {We perform this deconvolution with a `clean' algorithm analogous to that used in aperture synthesis imaging \citep[see][for a review]{cor99}.}

\begin{figure*}
\plotone{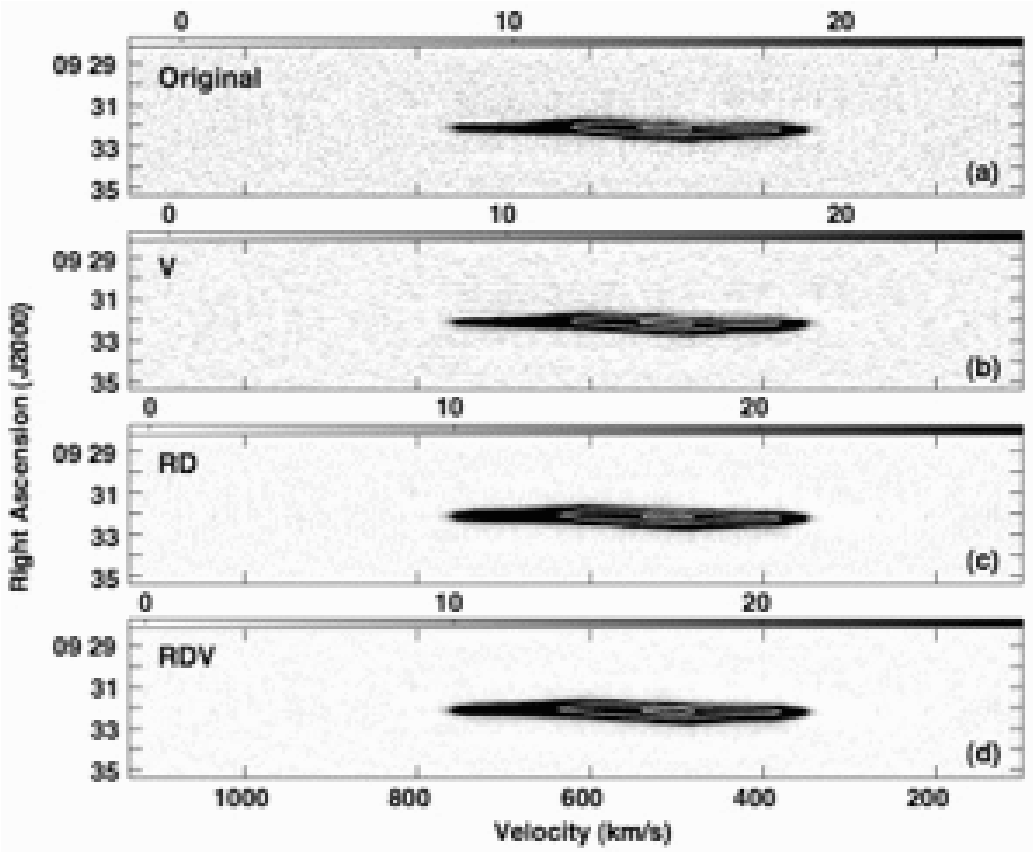}
\caption{{R.A. -- V plots at} Decl. = 21$^\circ$ 30$^\prime$ 03$^{\prime\prime}$, for the
Original cube (a), the V-smoothed cube (b), the RD-smoothed
cube (c), and the RDV-smoothed cube
(d) (see Table~\ref{table4}).  {V} is indicated at the bottom
of the lowest frame.  In each case, the greyscale range, marked
at the top of each frame, goes from
-1$\sigma$ to 10\% of the maximum in the frame, the latter being
253.0 mJy beam$^{-1}$ for (a), 
252.1 mJy beam$^{-1}$ for (b),
285.9 mJy beam$^{-1}$ for (c) and
284.5 mJy beam$^{-1}$ for (d).
Contours
at 50\% and 90\% of the peak are also shown.
\label{vslices}}
\end{figure*}

\begin{figure}
\plotone{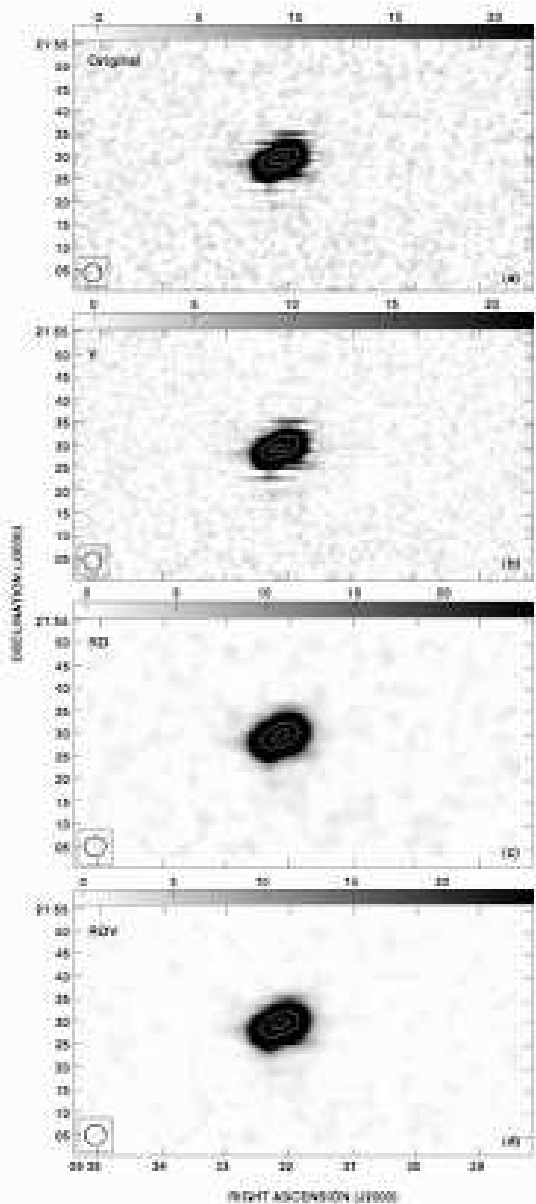}
\caption{{R.A. -- Decl. plots at}
V = 555.0 km s$^{-1}$, for the
Original cube (a), the V-smoothed cube (b), the RD-smoothed
cube (c), and the RDV-smoothed cube
(d) (see Table~\ref{table4}).  {RA} is indicated at the bottom
of the lowest frame.  In each case, the greyscale range, marked
at the top of each frame, goes from
-1$\sigma$ to 10\% of the maximum in the frame, the latter being
216.6 mJy beam$^{-1}$ for (a), 
218.0 mJy beam$^{-1}$ for (b),
247.1 mJy beam$^{-1}$ for (c) and
247.5 mJy beam$^{-1}$ for (d).
Contours
at 50\% and 90\% of the peak are also shown, as are the beam sizes
in each frame.
\label{rdslices}}
\end{figure}

{We use} an image-plane IDL-based implementation of the clean
algorithm written by Buie (2008)\nocite{MB}, modified by us  
to account for the multiple 
beams\footnote{See mbmclean.pro on our website.}.
Since the dominant contributing beam 
 was recorded in the 
grid data structure for each point, we were able to calculate 
the contribution to that point from a point source anywhere 
in the R.A. - decl. map at each velocity.
Each iteration consisted of identifying the strongest remaining 
emission in the map, then using the appropriate known beams 
 to remove the contributions of that point to the entire map. 
Iterations continued until the first negative clean component was reached,
or until the clean component reached the level of the noise 
which we took to be 2 mJy.  This clean procedure was carried out {at each azimuth, producing
12 cleaned NGC~2903 datacubes.} 


The cleaned cubes  
were 
then read into {AIPS}
for further reduction and analysis.  
Each of these cubes was inspected individually and some minor editing
(e.g. of remaining wandering birdie spikes farther from the galaxy emission)
was carried out.  All cubes at different azimuths
 were then averaged to form a single cube.  
 A subset of velocity-space in the cube was
then extracted so as to avoid noisy end channel{s} as well as contaminating
Galactic emission on the low velocity side.  The resulting
full-resolution cube will be designated as
the `Original' cube.
This cube was then smoothed, {in}
velocity alone (denoted V-{s}moothed){,} spatially alone (RD-{s}moothed)
and {both spectrally} and spatially 
(RDV-{s}moothed).  The spatial smoothing, in particular, 
ameliorates the striping
issue discussed in Sect.~\ref{galaxy_calibration}, improving the 
 the rms noise in the maps (see next section).
Finally, residual curvature in the
baseline was removed, point by point\footnote{The AIPS task, 
{\sc xbasl}, was
used.}. {The parameters of the final cubes are given in Table~\ref{table4}, and the cubes themselves may be obtained from our website.}

\begin{table*}[ht]
\begin{center}
\caption{Parameters of Cubes\label{table4}}
\begin{tabular}{lcccc}
\tableline\tableline
Parameter & Original & V-{s}moothed & RD-{s}moothed & RDV-{s}moothed \\
\tableline
Final velocity coverage 
(km s$^{-1}$)\tablenotemark{a} & 100.0 $\to$ 1132.8 
& 100.0 $\to$ 1132.8 & 100.0 $\to$ 1132.8 &100.0 $\to$ 1132.8\\
Channel width (km s$^{-1}$)& 1.293 & 1.293 & 1.293 & 1.293 \\
Velocity resolution, $\delta\,V$ (km s$^{-1}$)\tablenotemark{b} 
& 2.59 & 5.17 & 2.59 & 5.17\\
Spatial resolution, $\theta$ (arcsec)\tablenotemark{c}
& 234 & 234 & 270 & 270 \\
$\sigma$ (mJy beam$^{-1}$)\tablenotemark{d} 
& 1.1 & 0.80 & 0.61 & 0.44 \\
mean (mJy beam$^{-1}$)\tablenotemark{e} 
& 0.0069 & 0.0069 & 0.00988 & 0.00983 \\
M$_{HI~lim}$ (M$_\odot$)\tablenotemark{f} 
& 3.2$\,\times\,$10$^5$ & 4.7$\,\times\,$10$^5$ & 
1.8$\,\times\,$10$^5$ & 2.6$\,\times\,$10$^5$ \\
N$_{HI~lim}$ (cm$^{-2}$)\tablenotemark{g} 
& 3.5$\,\times\,$10$^{17}$ & 
5.1$\,\times\,$10$^{17}$ & 1.5$\,\times\,$10$^{17}$ &
2.1$\,\times\,$10$^{17}$\\
Max S/N\tablenotemark{h} 
& 533 & 730 & 1076 & 1475 \\
\tableline
\end{tabular}
\tablenotetext{a}{Bandwidth range  
after removing end channels and Galactic emission.}
\tablenotetext{b}{Since Hanning smoothing was applied, the
original velocity resolution is not equivalent to the channel width.}
\tablenotetext{c}{Full width at half maximum of the gaussian beam.}
\tablenotetext{d}{Rms noise over all regions
of the cubes in which the galaxy emission had been blanked.}
\tablenotetext{e}{Mean over all regions
of the cubes in which the galaxy emission had been blanked.}
\tablenotetext{f}{\HI mass limit ($3\,\sigma\,2\delta\,V$, as
described in Sect.~\ref{noise}).}
\tablenotetext{g}{\HI column density limit
($3\,\sigma\,2\delta\,V$, as
described in Sect.~\ref{noise}).}
\tablenotetext{h}{Maximum signal-to-noise (S/N) ratio of each cube.  
The S/N cubes are
described in Sect.~\ref{noise}.}
\end{center}
\end{table*}

\section{Detection Limits and Data Quality}
\label{data_quality}

{A selection of R.A. - V plots and R.A. - decl. plots {of the final cubes} 
are shown in Figs.~\ref{vslices} and \ref{rdslices}, respectively. The greyscale in the plots emphasizes low-intensity emission} { to illustrate the data sensitivity in low dynamic range regions of the cube as well as residual map errors near NGC~2903. We discuss these map properties in turn below.  }


\subsection{{Data Sensitivity}}
\label{noise}

{The mean, \sbar, and root-mean-square (RMS) noise, $\sigma$, of all regions {beyond the extended envelope of NGC~2903 (see Sect.~\ref{ngc2903_results})} in each cube is listed in Table~\ref{table4}. {As expected} {in these low dynamic range regions},  the final baseline 
is {consistent with zero ($\sbar << \sigma$) and a histogram of $\sigma$} over all line-free channels {is} gaussian.


{The variation in $\sigma$ as a function of R.A., decl. and V was examined. While we find that $\sigma$ is independent of R.A. and V, it does vary with declination,} 
a result that is illustrated in
Fig.~\ref{noise_fig}{a}.  
{This is primarily caused by the higher gain of the central beam, Beam 0
($\sim 11$K/Jy) relative to the outer ones ($\sim 8.5$ K/Jy).
The result is that declinations surveyed with the former have lower $\sigma$. 
The correspondence between the decl. of Beam 0 
for each drift (location of points along x axis of Fig.~\ref{noise_fig}d) 
and the minima in Fig.~\ref{noise_fig}a illustrates this effect. 
Different numbers of drift scans at some declinations (Fig.~\ref{noise_fig}d), 
uncertainties in beam calibration and variations in beam spacing 
(see Sect.~\ref{galaxy_mapping}) also contribute to changes in 
$\sigma$ with declination.}

From the noise values and some assumptions, we can compute 
detection limits for each cube in the low dynamic range regime.
We consider the limiting flux
integral, ${\cal S}_{lim}$, to be
from a{n unresolved}  signal that is at a 
 $3\,\sigma$ level   {in} 2
independent {channels}
{,}
\begin{equation}\label{Slim}
{\cal S}_{lim}\,=\,3\,\sigma\,2\,\delta\,V\,~~~{\rm{Jy\,km\,s}^{-1}}
\end{equation}
where {$\delta\,V$ (km s$^{-1}$) is the velocity
resolution (see Table~\ref{table4}) and $\sigma$ is in Jy beam$^{-1}$.}
The {minimum detectable \HI mass} is then,
\begin{equation}\label{mass_eqn}
{\rm M}_{HI~{lim}}\,=\,2.356\,\times\,10^5\,D^2\,{\cal S}_{lim}
\,~~~{\rm M}_\odot
\end{equation}
where $D$ is the distance (Mpc), {and the minimum detectable column density for a signal that
uniformly fills the beam is:} 
\begin{equation}\label{column_density_eqn}
{\rm N}_{HI~{lim}}\,=\,\frac{2.228\,\times\,10^{24}}
{(\theta\,\nu_c)^2}\,{\cal S}_{lim}\,~~~{\rm cm}^{-2}
\end{equation} 
where $\theta$ is the spatial resolution (arcsec, Table~\ref{table4}) and
$\nu_c$ is the central frequency  (GHz, Table~\ref{table3}).  
If the signal does not
uniformly fill the beam, then the right hand side of
Eqn.~\ref{column_density_eqn} must
be divided by an areal filling factor.

Plots of ${\rm M}_{HI~{lim}}$ and ${\rm N}_{HI~{lim}}$ as a function of {decl.} 
are shown in Figs.~\ref{noise_fig}b~and~\ref{noise_fig}c, respectively, 
and the mean values for each cube are given in Table~\ref{table4}.  
Note that the cubes smoothed in velocity have higher ${\rm M}_{HI~{lim}}$ and 
${\rm N}_{HI~{lim}}$ than their full resolution counterparts
because of the larger $\delta\,V$ of the former.  Note also that these results
are simply detection limits for the data, without imposing assumptions
about the properties of any companions that might be present.  The limits
shown in
Table~\ref{table4} are very low; for example, the column density limits
are lower than those of THINGS by two orders of magnitude (see
Footnote~\ref{things_footnote}).



\subsection{Residual Map Errors {near NGC~2903}}
\label{residual}

\begin{figure*}
\plotone{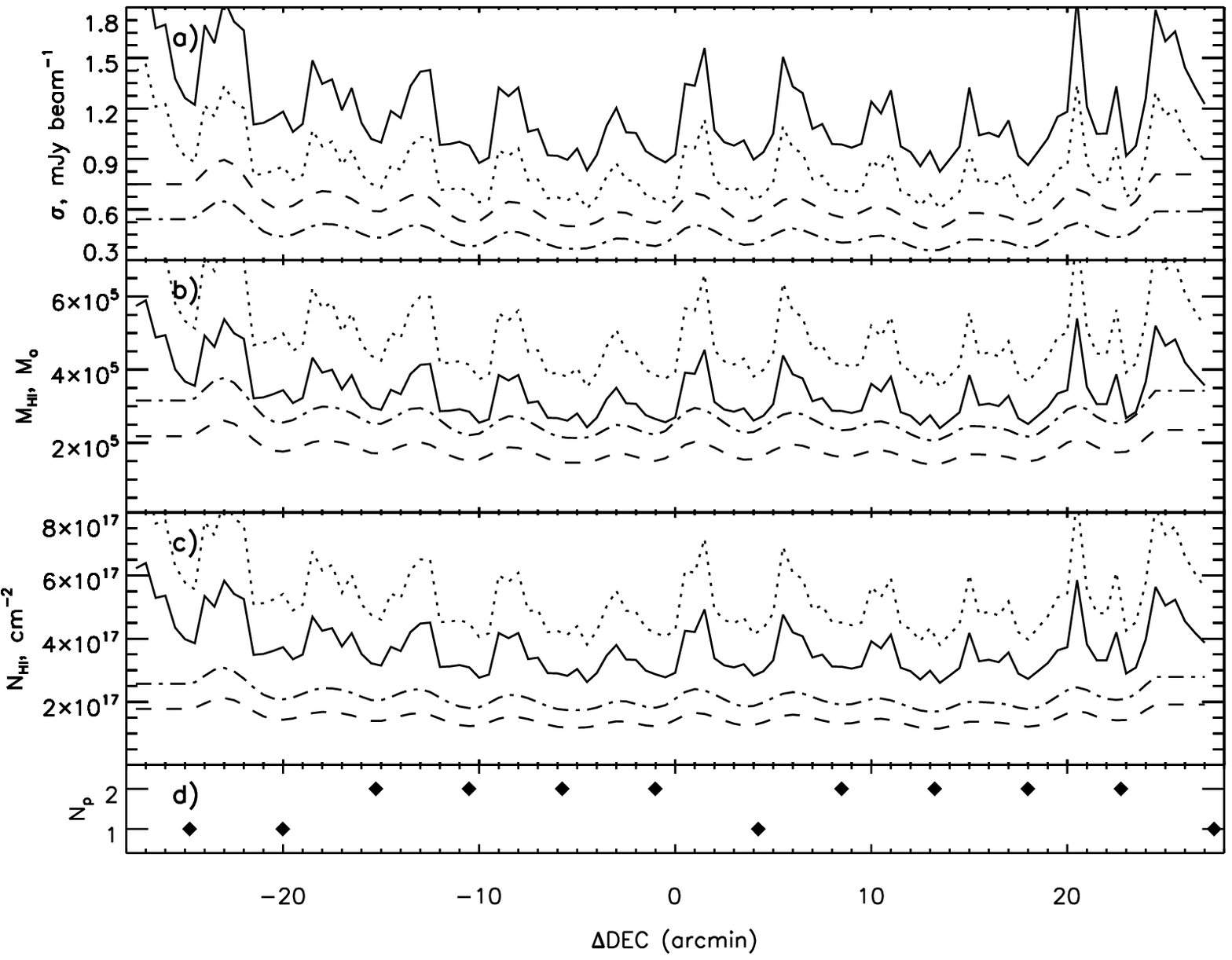} 
\caption{
Noise and detection thresholds as a function of {Dec}. $\Delta$ DEC
is relative to the optical center of NGC~2903 
(Table~\ref{table1}).
In each panel, the solid curve represents the original cube, the dotted
curve represents the V-{s}moothed cube, the dashed curve represents the
RD-{s}moothed cube, and the dash-dotted curve represents the RDV-{s}moothed
cube (Sect.~\ref{cleaning}, Table~\ref{table4}). 
(a) RMS noise $\sigma$. (b) Minimum detectable 
H\,I mass  $\mathrm{M} _{HI~lim}$, given by Eqn.
2. (c) {Minimum detectable} column density $\mathrm{N} _{HI~lim}$, given by 
Eqn.~\ref{column_density_eqn}.
 (d) Number of passes (i.e. drifts)
$N_p$ as a function of the declination of the central ALFA beam (Beam 0).
\label{noise_fig}}
\end{figure*}

{{Close inspection of} Figs.~\ref{vslices} and~\ref{rdslices} reveal residual map errors {near} NGC~2903 that remain even after the data reduction procedure discussed in Sect.~\ref{data_reduction}. These artifacts, described
below,
limit the dynamic range of the data
in regions occupied by emission from NGC~2903 itself, having a greater
relative effect  %
 near the `edges' of this emission.


One {artifact} that is evident in 
Figs.~\ref{vslices}a and{~\ref{vslices}}b
 is a faint ridge of emission seen on the {low-RA}
side of the galaxy emission, running close to and
parallel with the `edge' of the main galaxy emission.  
This ridge is due to imperfectly cleaned sidelobes (see Sect.~\ref{beam_reduction}). 
It is most evident in the data at full spatial resolution, and occurs
at typically a 2 to 5\% level in comparison to 
the} brightest galactic emission at the {same} velocity. 

{Another type of artifact is evident in 
Figs.~\ref{rdslices}a and ~\ref{rdslices}b, and can be attributed to residual striping due to scan calibration uncertainties 
(Sect.~\ref{galaxy_calibration}).}  These residual 
errors vary in strength but are typically at the level of
a few percent of the peak in any given channel
near the {edge} of the main
emission. They  produce the 
{`scalloping' of the edges of the
\HI distribution in NGC~2903 at low column densities} 
(e.g. Fig.~\ref{snmaps}).

Finally, the {second sidelobes} (outer coma lobes)
of the beam{s} were not {fully mapped in declination}, thus
limiting the effectiveness of the cleaning {in this dimension 
(see Sect.~\ref{beam_reduction})}. From an examination of the 
rather complex outer 
{lobes} observed in {R.A.}, we estimate that 
{second sidelobe} peaks at 
approximately 0.6\% of the central beam peak 
could contribute to emission as far
as 10$^{\prime}$ away in {decl.}
(note, however, that {this varies}  beam to beam and azimuth to azimuth).

{Considering these residual map errors and typical peak \HI flux densities measured for NGC~2903 in our data, we caution against simple interpretations of emission below 
a few percent in any given channel over the emission region occupied by
NGC~2903 itself.
We estimate that a more conservative detection criterion of 
$\sim 15\sigma2\delta V = 5\,{\cal S}_{lim}$ 
($\approx$ 10$^{18}$ cm$^{-2}$ in column density) is likely appropriate 
in these high-dynamic range regions; that is, the detection limits
of Table~\ref{table4} should be increased by a factor of 5 for regions
occupied by NGC~2903 emission itself.  We
defer the detailed 
analysis required to determine the outer 
\HI morphology of NGC~2903 to a future paper.  
We emphasize that these artifacts arise only near bright emission, 
and thus the sensitivity limits given in Table~\ref{table4} using
 ${\cal S}_{lim}$ in Eqn.~\ref{Slim} are appropriate for regions of the
data cubes in which we search for companions (see Sect.~\ref{companions}). }



\section{Results}
\label{results}

\subsection{{Basic Properties of} NGC~2903}
\label{ngc2903_results}


 {While a detailed analysis of the \HI morphology and kinematics of NGC~2903 is beyond the scope of this paper, we present some of its basic properties here to illustrate the content and quality of our datacubes.}

\subsubsection{Global Parameters}
\label{globalparms}

 Fig.~\ref{ngc2903_globalprofile} shows the global profile of NGC~2903,
and corresponding global parameters are given in Table~\ref{table5}.
The profile shape agrees well with 
previously published plots \citep{won06, hew83} 
and our integrated flux
density (Table~\ref{table5}) 
agrees with the result of Braun et al. (2007)\nocite{bra07}
to within errors. 
 There is an obvious asymmetry in the galaxy, such 
that the low velocity peak (north-east side of galaxy)
 is higher than the high velocity peak (south-west side).  The integrated
flux on the low velocity side of the galaxy is 13\% higher than on
the high velocity side, denoting an intrinsic asymmetry in the \HI
distribution
of that order.

\subsubsection{{Morphology and Kinematics}}
\label{envelope}

We present the integrated intensity and intensity-weighted mean velocity
fields, from the RDV-smoothed cubes, in Fig.~\ref{momnts}. 
{A small,} previously unknown \HI companion can be seen 24.8$^\prime$ 
(64.3 kpc in projection) to the north-west.  The
eastern companion, UGC~5086, is enveloped in the \HI emission
from NGC~2903.  \HI companions will be discussed
further in Sect.~\ref{companions}.  


 We find a very large \HI envelope around NGC~2903, even accounting for the Arecibo beam and residual map errors.  For comparison, the grey contour 
in Fig.~\ref{momnts}a shows the outermost significant integrated 
intensity level (1$\,\times\,$10$^{19}$ cm$^{-2}$) in the archived 
D-configuration VLA observations (see Sect.~\ref{ngc2903}), smoothed 
to the same resolution as the Arecibo data in the figure. The major 
axis diameter at 10$^{18}$ cm$^{-2}$ in our Arecibo data -- which 
should be immune to residual map errors 
(see Sect.~\ref{residual}) -- is $d_{HI}=40.7^\prime$ (105 kpc) 
after correcting for the beam, nearly twice the value measured from the 
VLA data.   Thus, the \HI extent of NGC~2903 is at least 3.2 times its 
optical diameter (Table~\ref{table1}) and ranks among the largest 
known \citep{mat01,del04,spe06,oos07,cur08}.

 The velocity field of NGC~2903
(Fig.~\ref{momnts}b) shows regular rotation,
with the north-east side advancing with respect to the center.
The contours indicate that the {outer} \HI disk of the galaxy is warped
in spite of its apparent isolation.  
A position-velocity plot along a 270$^{\prime\prime}$ wide strip
of the major axis is shown
in Fig.~\ref{majax}. 
{The inner rotation curve of NGC~2903 appears to be regular, 
but it is strongly biased by beam smearing and not a good indicator 
of the gravitational potential in these regions. 
We find no evidence for gas at anomalous velocities 
at the sensitivity and resolution of our data.}

\begin{table}[ht]
\begin{center}
\caption{\HI Properties of NGC~2903\label{table5}}
\begin{tabular}{lc}
\tableline\tableline
Parameter & Value \\
\tableline
$\Delta\,$V$_{50}$ (km s$^{-1}$)\tablenotemark{a}& 370 $\pm$ 2 \\
$\Delta\,$V$_{20}$ (km s$^{-1}$)\tablenotemark{b}& 383 $\pm$ 2\\
V$_{sys}$ (km s$^{-1}$)\tablenotemark{c} & 555 $\pm$ 2 \\
$\int\,S_V\,dV$ (Jy km s$^{-1}$)\tablenotemark{d}
& 255 $\pm$ 15 \\
M$_{HI}$ (\msun)\tablenotemark{e} 
& (4.8 $\pm$ 0.3)$\,\times\,$10$^9$\\
$d_{HI}/d_{opt}$\tablenotemark{f} & 3.2\\
\tableline
\end{tabular}
\tablenotetext{a}{Full width at 50\% of the average {of the 2 profile peaks.}}
\tablenotetext{b}{As in {\it a} but at 20\%.}
\tablenotetext{c}{{V at the midpoint of $\Delta V_{20}$.}} 
\tablenotetext{d}{Integral of flux density
{associated with  NGC~2903.  The \\dominant uncertainty is from the baseline flattening; \\excluding this effect, the uncertainties are of order 1\%.}}
\tablenotetext{e}{\HI mass, {from Eqn.~\ref{mass_eqn} substituting $\int S_V dV$ for ${\cal S}_{lim}$.}}
\tablenotetext{f}{{ratio of \HI to optical major axis diameter. $d_{HI}$ is \\measured at $10^{18}\,$cm$^{-2}$ and $d_{opt}$ is from Table~\ref{table2}.}} 
\end{center}
\end{table}

\subsection{\HI Companions of NGC~2903}
\label{companions}

\subsubsection{The Known Companion, UGC~5086}
\label{ugc5086_sec}
{UGC~5086 is the only 
previously
known companion of NGC~2903 that lies} within the surveyed region (see Sect.~\ref{ngc2903}).
  This galaxy overlaps the large
\HI envelope of NGC~2903 in both position and velocity space.
Fig.~\ref{ugc5086}, which shows a single channel at the systemic velocity
of UGC~5086, illustrates
this blending.  
 A global profile from a 
3$^\prime$ $\times$ 3$^\prime$ box centred on UGC~5086 in the RDV-smoothed cubes 
 yields a spectrum (not shown) with a peak of 
4.4 mJy {at V$ = 497\,$km s$^{-1}$}, resulting in a measured \HI mass
of  M$_{HI}=\,9.0 \times 10^{6}$ M$_\odot$ (Eqn.\ref{mass_eqn}
substituting the integrated flux of the profile for ${\cal S}_{lim}$).
{While the velocity of the spectral peak agrees with the systemic velocity
 of UGC~5086 
(Table~\ref{table2})}, 
the detected signal could {also arise from the envelope 
of NGC~2903 itself. To help distinguish between
signal from UGC~5086 and NGC~2903, we have searched 
our reduced VLA D-configuration (Sect.~\ref{ngc2903}) data, 
which clearly separate the two galaxies spatially.  
We find no emission from UGC~5086, and place an upper 
limit on the \HI mass in a single 54\arcsec\ beam of 
M$_{HI~lim} = 5.6 \times 10^5\,$M$_{\odot}$ from the VLA data. Thus, our
Arecibo detection is primarily from NGC~2903 iself.}

\begin{figure}
\epsscale{0.8}
\plotone{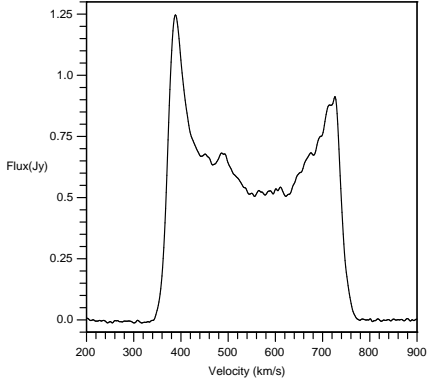} 
\caption{{Global} profile of NGC~2903. 
Errors (ignoring any uncertainty in baseline flattening) 
are typically within 1\% except where the profile approaches 
zero.
\label{ngc2903_globalprofile}}
\end{figure} 

\begin{figure*}
\plotone{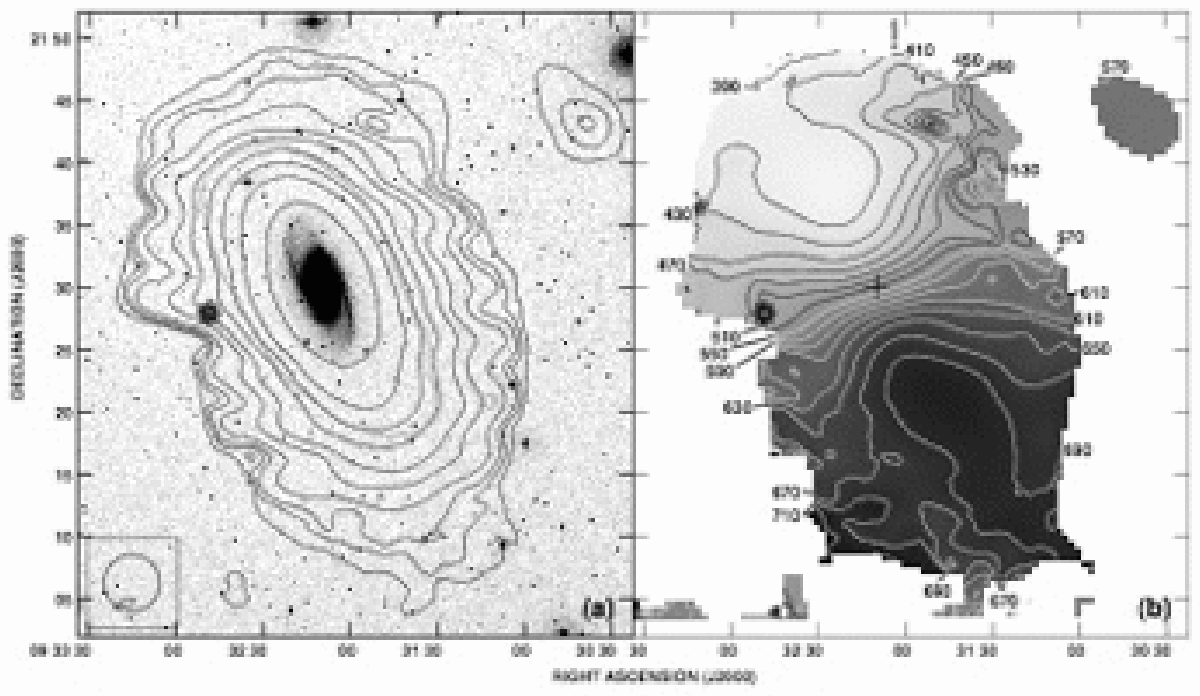}
\caption{Moment maps of NGC~2903, constructed from the RDV-smoothed
cubes.  
The location of the galaxy, UGC~5086 
is marked with a star. {Detailed maps of the companion to the north-west of NGC~2903 are in Fig.~\ref{companion_fig}.}
(a) Total intensity H\,I map over the DSS2 Blue image, the latter
shown in an arbitrary greyscale.  Contours are at
0.02, 0.06, 0.10, 0.20, 0.50, 1.0, 2.5, 5, 10, and 25
Jy beam$^{-1}$ km s$^{-1}$. The peak is 68.3 Jy beam$^{-1}$ km s$^{-1}$.
The beam is shown at lower left. 
A conversion to column density requires
a multiplication 
by 1.52$\,\times\,$10$^{19}$
cm$^{-2}$ (Jy beam$^{-1}$ km s$^{-1}$)$^{-1}$.
{Note that there may be residual map errors near NGC~2903 below $\sim 10^{18}\,$cm$^{-2}$; see Sect.~\ref{residual}.}
The grey curve shows the outermost 
significant integrated intensity level (1$\,\times\,$10$^{19}$ cm$^{-2}$) in the archived D-configuration VLA observations (see Sect.~\ref{ngc2903}), 
after smoothing the VLA cube to the same 
spatial resolution as the Arecibo data.
(b) Intensity-weighted mean velocity contours over a greyscale 
from the same image.
Contours, in km s$^{-1}$, are labelled and are spaced 20 km s$^{-1}$
apart.  The optical center of the galaxy is marked with a cross.
\label{momnts}}
\end{figure*}

 UGC~5086 is well resolved in the SDSS 
Data Release 6 (DR6).  Its image
appears almost perfectly circular
and it has a red colour (B$_0$ - V$_0$ = 0.79, using SDSS magnitudes and
applying transformations referenced in Sect.~\ref{2mas}). Given the
absence of \HI and known optical parameters, it is likely that 
UGC~5086 is a dwarf
spheroidal galaxy without \HI content.  We note that near-UV emission
can be seen in 
UGC~5086 also\footnote{{\tt http://skyview.gsfc.nasa.gov}},
suggesting the presence of a young
stellar population.  This is similar to the dwarf spheroidal galaxy,
Fornax, which contains both old, intermediate, and young stars
\citep{bat06}.



\subsubsection{Search for New Companions}
\label{new_companions}


\begin{figure}
\epsscale{1.0}
\plotone{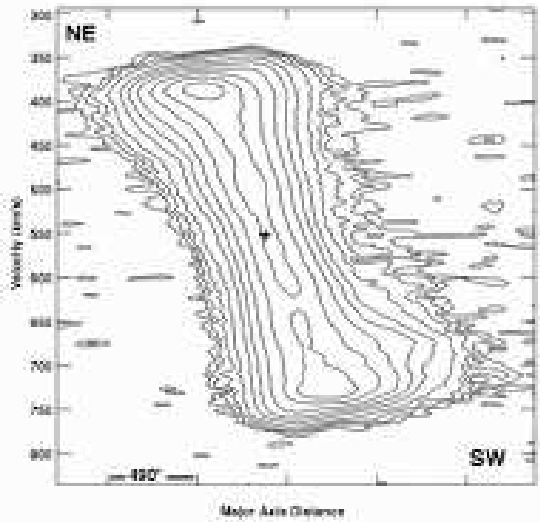}
\caption{Position-velocity plot along the major axis of NGC~2903, 
from the RDV-smoothed cube, averaged over
a width equivalent to the beam size.  The center is marked with a cross.
Contours are at 1.0 (2$\,\sigma$), 1.8, 3.0, 6.0, 15, 30, 60, 120, 200,
and 300 mJy beam$^{-1}$.  The north-east (NE) and south-west (SW) sides
of the galaxy are labelled and tickmarks along the position axis are
separated by 490$^{\prime\prime}$.   
{Note that there may be residual map errors below 
$\sim$ 3\% near the galaxy; see Sect.~\ref{residual}.}
\label{majax}}
\end{figure}

{A visual search of the datacubes has revealed a new \HI-rich companion to NGC~2903}, visible in Fig.~\ref{momnts}. This companion, which we designate N2903-HI-1, will be discussed further in 
Sect.~\ref{2mas}.  
In order to detect companions in a more quantitative fashion,
it is important {to account for the variation in map noise, $\sigma$, with 
decl., illustrated in Fig.~\ref{noise_fig}a. 
To this end, we formed S/N cubes\footnote{The S/N cubes may be download from our website.} by dividing each datapoint by the $\sigma$ corresponding to its decl., 
and searched for emission exceeding ${\cal S}_{lim}$ from Eqn.~\ref{Slim}.}
{Specifically}, we integrated each S/N cube over all {V}
including only those points that exceed $3\,\sigma$ over at least
two adjacent independent velocity resolution elements\footnote{Our
AIPS-compatible routine (xsmc), written for this purpose,
is available on our website.}.  The resulting summed maps, which
emphasize {faint emission}, are shown in
Fig.~\ref{snmaps}.

Fig.~\ref{snmaps} confirms that N2903-HI-1
{is the only {\it bona fide} \HI-rich companion to NGC~2903 in our data}. {While other isolated non-zero}
 pixels are also seen in the various maps, {they are clearly random noise peaks.} 
 {This is corroborated by the lack of correlation between the locations of these pixels in the different maps.}

There is a possibility that companions may lie within the spatial
region over which NGC~2903 is found, but at anomalous velocities
in comparison to NGC~2903.  We have searched through this parameter space {and} find no evidence for such \HI clouds
(see also the major axis slice of Fig.~\ref{majax}).

{It is also plausible that} our search has missed companions which overlap NGC~2903 in {\it both} position and velocity space, {a possibility raised by the location of UGC~5086 (Sect.~\ref{ugc5086_sec}).  The \HI disk of NGC~2903 occupies an exceedingly small percentage of the total volume surveyed, however, making such a coincidence highly unlikely.} Moreover,  if  {a starless} \HI emission feature 
exists within the \HI position-velocity envelope 
of NGC~2903, then it becomes moot as to whether such a feature should
simply be considered part of NGC~2903 itself.


\begin{figure}
\plotone{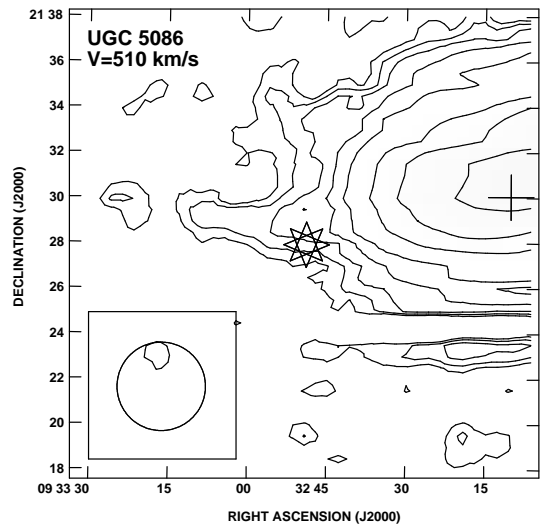} 
\caption{A single channel of the V-smoothed cube 
centered on the position and velocity (Table~\ref{table2}) of
the companion, UGC~5086.  The companion is seen within the region
of a protrusion to the east of the H\,I envelope associated with
NGC~2903.  UGC~5086 is marked with a star and the
center of NGC~2903 is marked with a cross to the west.  Contours are
at 1.6 (2$\,\sigma$), 3.0, 5.0, 10, 20, 50, 100, and 200 mJy beam$^{-1}$. 
Note that there may be residual map errors below 
$\sim$ 3\%; see Sect.~\ref{residual}.  The beam is shown at lower left.
\label{ugc5086}}
\end{figure}

\subsubsection{The New Companion, N2903-HI-1}
\label{2mas}

{N2903-HI-1 is separated from NGC~2903} by 24.8\arcmin\ spatially (64.3 kpc in projection)  and by +26 km s$^{-1}$ in
velocity (cf. Tables~\ref{table1} \& \ref{table6}).  
{Its} global profile is shown in 
Fig.~\ref{companion_globalprofile}, 
and related parameters are given in Table~\ref{table6}.  We also show the
total intensity map, a position-velocity slice, and the 1st and
2nd moments {of the \HI distribution} in Fig.~\ref{companion_fig}.  The full velocity resolution
cube has been used for the latter two maps since the profile is narrow.



Fig.~\ref{companion_fig}a illustrates that N2903-HI-1 is elongated north-east to south-west and has a 
`cometary' or `head-tail' morphology.
In spite of the large beam size, the companion is
 spatially resolved in all cubes, hence its radius, $R$,
 can be measured.
This has been done via a
gaussian fit to the highest spatial resolution data and
deconvolving the beam (Table~\ref{table6}).
 
We {do not} find 
convincing evidence of systematic
motion in our data, either from the major axis slice 
(Fig.~\ref{companion_fig}b), the 1st moment map
(Fig.~\ref{companion_fig}c), or channel maps (not shown).  
 A rotating disk {with an inclination $i >15^\circ$ would exhibit a} 
gradient across the disk
that is greater than {the} typical velocity dispersion of 6 km s$^{-1}$ in Fig.~\ref{companion_fig}d. 
 Thus, if the \HI represents a disk in rotation, we might have expected, given our fine velocity resolution, to have seen some 
evidence for this. Given the elongated
morphology of the \ion{H}{1},  it seems
more likely that it is being either tidally perturbed or
ram pressure stripped via passage through a gaseous medium.  
Head-tail morphologies
are typically seen in the latter case, but because of low spatial
resolution, the former cannot be ruled out.

\begin{figure*}
\epsscale{0.9}
\plotone{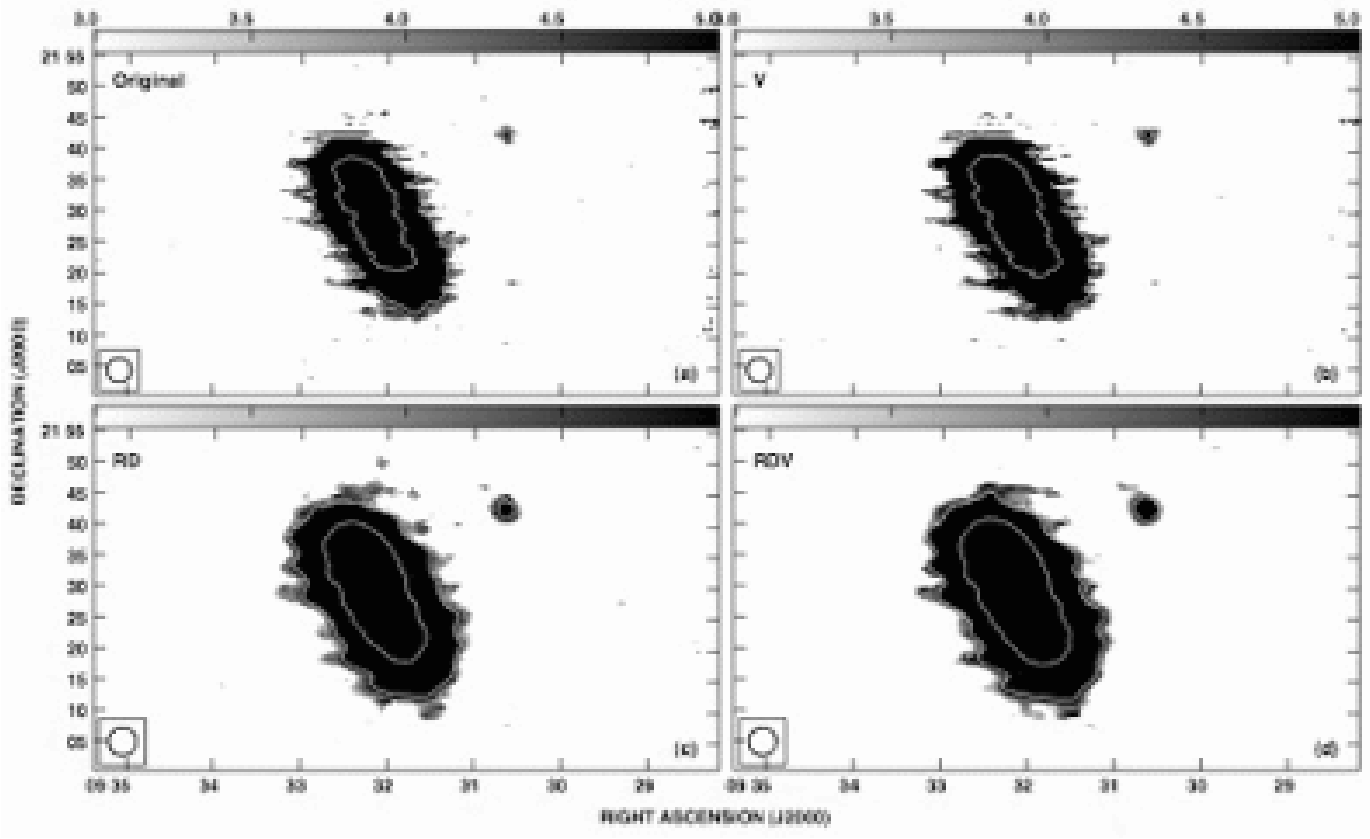}
\caption{{Summed} S/N {maps} for each of the four cubes (labelled at top
left in each frame). Only points  $\ge\,3\,\sigma$
that are detected in $\ge\,2\,\delta\,V$ independent velocity resolution
elements, as described in Sect.~\ref{new_companions}, 
have been included in the maps.
Low levels have been emphasized, with a greyscale range 
($3\sigma - 5\sigma$) shown at the top, and  
the $5\sigma$ and $50\sigma$ contours are also shown.
The beam is shown at lower left.
Peak levels are $161\sigma$,  $216\sigma$,  $256\sigma$, and 
 $343\sigma$ for (a), (b), (c), and (d), respectively. {Note that there may be residual map errors below the $5\sigma$ contour in NGC~2903; see Sect.~\ref{residual}.}
\label{snmaps}}
\end{figure*}


To determine a total dynamical mass, ideally we would want to associate
the radius of N2903-HI-1 with some rotational velocity.  As we do not
know the precise geometry of the system and do not see rotation, 
this association cannot be made.  However,
the measured line width and radius should provide us with some measure
of the total mass which we now estimate under two 
 assumptions that should encompass the extrema of possibilities
(see Westmeier et al. 2005a\nocite{wes05a} for a similar approach).

 First, in the event that
a rotation might have remained undetected
(for example, if the companion represents a rotating galaxy 
that is near face-on),
 the dynamical mass, ${\rm M}_{dyn}$, can be estimated via,
\begin{equation}
\label{mdyn}
{\rm M}_{dyn}\,\sin^2 i\,=\,\frac{R}{G}\,\left(\frac{\Delta V_{50}}{2}\right)^2
\end{equation}
{where $\Delta V_{50}$ is
 the width of the profile in Fig.~\ref{companion_globalprofile} at 50\% of the peak (Table~\ref{table6})}. $\Delta V_{50}$ 
has been adopted, rather than $\Delta V_{20}$, 
since the latter is a measurement
at a level within the noise.  With the assumption of rotation,
$R$ may be overestimated
if there is a cometary tail that does not take part in the rotation; however,
the velocity width, which has been minimized by the choice of
 $\Delta V_{50}$,  has a greater effect.   Together with what would
be a substantial correction
for the unknown inclination, the 
result for M$_{dyn}$ (Table~\ref{table6})
will be a minimum.

Secondly, 
for comparative purposes,
 we assume virial equilibrium for which,
\begin{equation}
\label{mvir}
{\rm M}_{vir}\,=\,\frac{5\,R\,\left(\Delta V_{50}\right)^2}
{8\,G\,\ln{2}}
\end{equation}
Here, we have related the mean square velocity of the particles,
$\langle v^2 \rangle$, to the observed 
FWHM velocity width, $\Delta V$, according
to $\langle v^2 \rangle\,=\,3\,(\Delta V)^2/(8\,\,\ln{2})$ (Westmeier et al.
2005b\nocite{wes05b}).  

The results
(Table~\ref{table6}) indicate a total mass for N2903-HI-1 which exceeds
$10^8$ M$_\odot$ by either estimate (Table~\ref{table6}),
and {we adopt} the mean, 
$3\,\times\,10^8\,$M$_\odot$, as a `characteristic' dynamical mass. 
Although the error bars are substantial,
it is nevertheless clear 
 that the \HI mass of N2903-HI-1 is only a small
fraction of its total mass.

\begin{table}[ht]
\begin{center}
\caption{Properties of N2903-HI-1\label{table6}}
\begin{tabular}{lc}
\tableline\tableline
Parameter{\tablenotemark{a}} & Value \\
\tableline
R.A. (h m s)\tablenotemark{b} & 09 30 38 \\
Decl. ($^\circ$ $^{\prime}$ $^{\prime\prime}$)\tablenotemark{b} & 21 43 08\\
$\Delta\,$V$_{50}$ (km s$^{-1}$)\tablenotemark{c}& 23.6 $\pm$ 5.2 \\
$\Delta\,$V$_{20}$ (km s$^{-1}$)\tablenotemark{d}& 40.2 $\pm$ 5.2\\
V$_{sys}$ (km s$^{-1}$)\tablenotemark{e} &  582 $\pm$ 4\\
$\int\,S_V\,dV$ (Jy km s$^{-1}$)\tablenotemark{f}
&  0.14 $\pm$ 0.02\\
M$_{HI}$ (\msun)\tablenotemark{g} 
& (2.6 $\pm$ 0.3) $\times$ 10$^6$\\
$R$ (arcsec)\tablenotemark{h} &  90$\,\pm\,$40 \\
~~~~~(kpc) & 3.9$\,\pm\,$1.7 \\
M$_{dyn}\,$sin$^2$(i) (M$_\odot$)\tablenotemark{i} & $(1.3 {\,\pm \,1.1})\,\times\,10^8$\\
M$_{vir}$  (M$_\odot$)\tablenotemark{i} & $(4.5 {\,\pm\, 4.0\,})\,\times\,10^8$\\
\tableline
\end{tabular}
\tablenotetext{a}{Table~\ref{table5} provides definitions when not indicated
here.}
\tablenotetext{b}{Position {of} peak 
of the {total intensity} map (Fig.~\ref{companion_fig}a).\\ 
The uncertainty is {$\sim15\arcsec$ (1/2 the cellsize) in RA} and \\ $\sim1\arcmin$ (1/4 of the beam) in decl.}
\tablenotetext{c}{Full width at 50\% of the {profile} peak.\\
The error bar encompasses variations between the cubes.}
\tablenotetext{d}{As in {\it b} but at 20\%.}
\tablenotetext{e}{{Average of V at the midpoint of $\Delta V_{20}$ and  $\Delta V_{50}$ .}} 
\tablenotetext{f}{{Uncertainty} is dominated by
variations between\\ the different cubes and choice of
velocity window.}
\tablenotetext{g}{Assuming the distance is the same as NGC~2903.}
\tablenotetext{h}{{Radius of N2903-HI-1}, from 1/2 of
the FWHM found from \\deconvolved gaussian fits 
to the total intensity map.}
\tablenotetext{i}{Dynamical (M$_{dyn}$), and virial
(M$_{vir}$) masses, as given
by\\ Eqns.~\ref{mdyn} and \ref{mvir}, respectively.}
\end{center}
\end{table}


{Does N2903-HI-1 have an optical counterpart?}
Because of the elongated shape of the \HI emission, {it is possible}
 that an optical galaxy could be displaced with respect
to the \HI central peak.  We have
therefore searched the {SDSS DR6} database  over the original full beam width, 
(3.9$^\prime$),
centered on the peak position of N2903-HI-1 (Table~\ref{table6}).
There are 207 {catalogued sources}  in this area,
each with photometric, rather than spectroscopic redshifts.
  Given that
the dispersion between spectroscopic and photometric redshifts
is of order $\approx\,$0.06 \cite{csa03} at low redshift, we have identified all
 galaxies with redshifts less than
0.06 over this spatial region.  The result gives 34 galaxies, all of
which are plotted in  Fig.~\ref{companion_fig}a.  
Stellar masses have been calculated for each of these galaxies
according to the formalism of Bell et al.
(2005), assuming a Kroupa (2001)\nocite{kro01}
 Initial Mass Function (IMF), and assuming that they are at the
distance of NGC~2903.  If a modified
Salpeter IMF is used instead \cite{bel03},
the stellar mass increases by only a factor of
1.4.  


The galaxy, SDSS
J093039.96+214324.7 
(see Table~\ref{table7}), which is marked in red in Fig.~\ref{companion_fig}a, 
is the most likely optical counterpart
{for several reasons}.
{First, t}he position of this galaxy agrees with
 the peak of N2903-HI-1 within errors (Tables~\ref{table6}, \ref{table7}). {It is also  
 significantly brighter (by at least $2.9\,$mag in $g$), larger and has more stellar mass}\footnote{The stellar mass of the second most massive galaxy is only 37\%
lower, but it is separated from the peak of N2903-HI-1 by 3.9 arcmin.} than other candidates in the field.
{A 2nd Digital Sky Survey (DSS2) image of this galaxy is
shown  in Fig.~\ref{dss2_optical_companion}}, and 
reveals a galaxy that is elongated roughly north-south, {similar to the \HI morphology of N2903-HI-1 (see Fig.~\ref{companion_fig}) but at a different position angle. }
Assuming that J093039.96+214324.7 is
 at the distance
{of} NGC~2903, we list its properties Table~\ref{table7}.

\begin{table}[ht]
\caption{Properties of {J093039.96+214324.7}\label{table7}}
\begin{tabular}{lc}
\tableline\tableline
Parameter & Value \\ 
\tableline
2a$\,\times\,$2b ($^{\prime\prime}\,\times\,^{\prime\prime}$)\tablenotemark{a} 
& 22 $\times$ 12 \\
~~~~~~~~~   (pc $\times$ pc)\tablenotemark{a} & 950 $\times$ 518 \\
c{\it z} (km s$^{-1}$)\tablenotemark{b} & 200 \\
u, g, r (mag)\tablenotemark{c} & 18.71, 18.14, 18.00 \\
B$_0$, V$_0$ (mag)\tablenotemark{d} & 18.28, 17.98 \\
$M_B$, $M_V$ (mag)\tablenotemark{e} & -11.47, -11.77 \\
M$_{\star}$ (M$_\odot$)\tablenotemark{f} & 1.8 $\times$ 10$^6$ \\
M$_{HI}$/L$_B$  (M$_\odot$/$L_\odot$)\tablenotemark{g} & 0.43 \\
\tableline
\end{tabular}
\tablenotetext{a}{{M}ajor 
$\times$ minor axis dimensions
from the 2$\,\sigma$ contour of \\
Fig.~\ref{dss2_optical_companion}.}
\tablenotetext{b}{Photometric 
redshift 
from the SDSS DR6.}
\tablenotetext{c}{Magnitudes in SDSS u, g and r bands.}
\tablenotetext{d}{B and V magnitudes from the
transformation \\of Lupton (2005)\nocite{lup05}, corrected for Galactic extinction \\
(Schlegel et al. 1998\nocite{sch98}) .}
\tablenotetext{e}{Absolute B and V magnitudes.}
\tablenotetext{f}{Stellar mass, {derived from ${\rm B}_0\,-\,{\rm V}_0$ and $M_V$}  
assuming \\a Kroupa IMF (Bell et al.\ 2005).}
\tablenotetext{g}{Ratio of \HI mass (Table~\ref{table6}) to blue luminosity,
the \\latter from $M_B$, with a Solar B-band magnitude of 5.48.}
\end{table}

At the distance of NGC~2903, the properties of
J093039.96+214324.7
 indicate that it is a low luminosity 
dwarf galaxy with a linear diameter of approximately 1 kpc.
If this galaxy is the optical counterpart of N2903-HI-1, then
its stellar mass  rivals
 its \HI mass but
 is at least two orders of magnitude less than its total mass,
leading to the conclusion that the system is dark matter dominated.
(If J093039.96+214324.7 is not the optical counterpart, the same
conclusion is reached.)
The
\HI envelope of N2903-HI-1 extends to $\sim 8$ optical radii.
 Comparing
 J093039.96+214324.7 to known Local Group dwarfs \cite{mat98} or
faint irregular galaxies \cite{beg08}, we find that its
colour,
 (B$_0$ - V$_0$ = 0.31), 
radius, line width,
 total mass, and \HI mass to blue light
ratio (M$_{HI}$/$L_B$) fall within observed ranges for these other known
systems.
Although its recessional velocity differs from that of N2903-HI-1,
the difference falls within typical errors for photometric redshifts
of nearby systems \cite{csa03}.  Thus, the position
and properties of 
J093039.96+214324.7, in comparison to the other galaxies in
the region, all suggest that this galaxy is the likely
optical counterpart of N2903-HI-1.  
A spectroscopic redshift for this system would decide the matter.

 We note that the \HI mass of N2903-HI-1 also falls within the range of
 Galactic HVCs for which there are
distance constraints \citep{put02, tho08, wak07, wak08}
and at the upper end of the \HI mass distribution found
for the HVCs around M~31 and M~33 \citep{thi04, wes05a, wes07}. {However, its} separation from
NGC~2903 (64 kpc in projection) is larger than the population of HVCs around
the Milky Way
(less than 10 - 15 kpc typically,
Thom et al. 2008\nocite{tho08}) or M~31 (within 50 kpc,
Westmeier et al. 2007\nocite{wes07}). In combination with the evidence for an optical counterpart, we consider this interpretation for N2903-HI-1 
to be less likely.

\section{Discussion}
\label{discussion}

\begin{figure}
\plotone{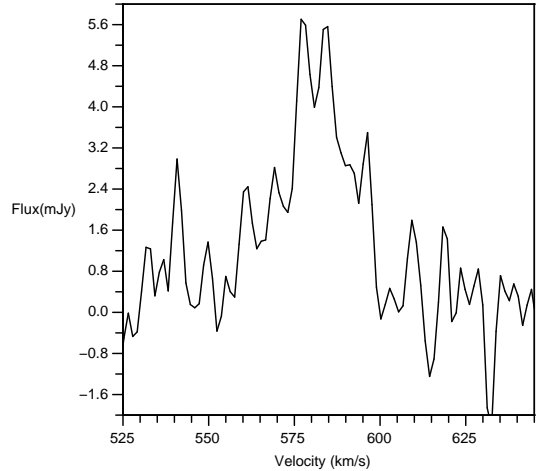} 
\caption{{G}lobal profile of the companion,
N2903-HI-1, obtained from
the RD-smoothed cubes. 
\label{companion_globalprofile}}
\end{figure}

In our targeted sensitive \HI survey of the isolated 
Milky-Way analogue, NGC~2903,
we have discovered one new companion, the \HI object, N2903-HI-1, 
which is dark-matter dominated (Table~\ref{table6}) and
 is likely associated
with the optical dwarf galaxy,
 SDSS J093039.96+214324.7. 
 New discoveries of dwarf satellites of the Milky Way 
from the SDSS place their typical total masses in the $10^{6-7}$ M$_\odot$
range \citep{sim07}
 which is lower than the total mass of $>$ 10$^8$ M$_\odot$ 
(Table~\ref{table6}) found for N2903-HI-1.
Combining data from previously known MW companions, the new SDSS companions,
 and those of M~31,
 most satellites which still contain 
detectable \HI lie beyond 300 kpc radius, the implication being that
\HI has been stripped in closer systems \citep{put08, grc08}, although notable
exceptions also occur (e.g. the Large and Small Magellanic Clouds).
N2903-HI-1, at a projected distance of 64 kpc from NGC~2903,
would have to lie at least 293 kpc in front of or behind
NGC~2903 to be at a true separation $>$ 300 kpc.  Although this is
possible, it is more likely that the companion is closer to NGC~2903,
yet has retained its \HI --- a
result that
 may be related to its high dynamical mass in comparison
to most Milky Way systems within the same radius.  Thus, if ram pressure
stripping is occurring, as suggested by the head-tail morphology
of N2903-HI-1, the process may be slower because of its high
 dynamical mass.
 
 \begin{figure*}
\epsscale{0.8}
\plotone{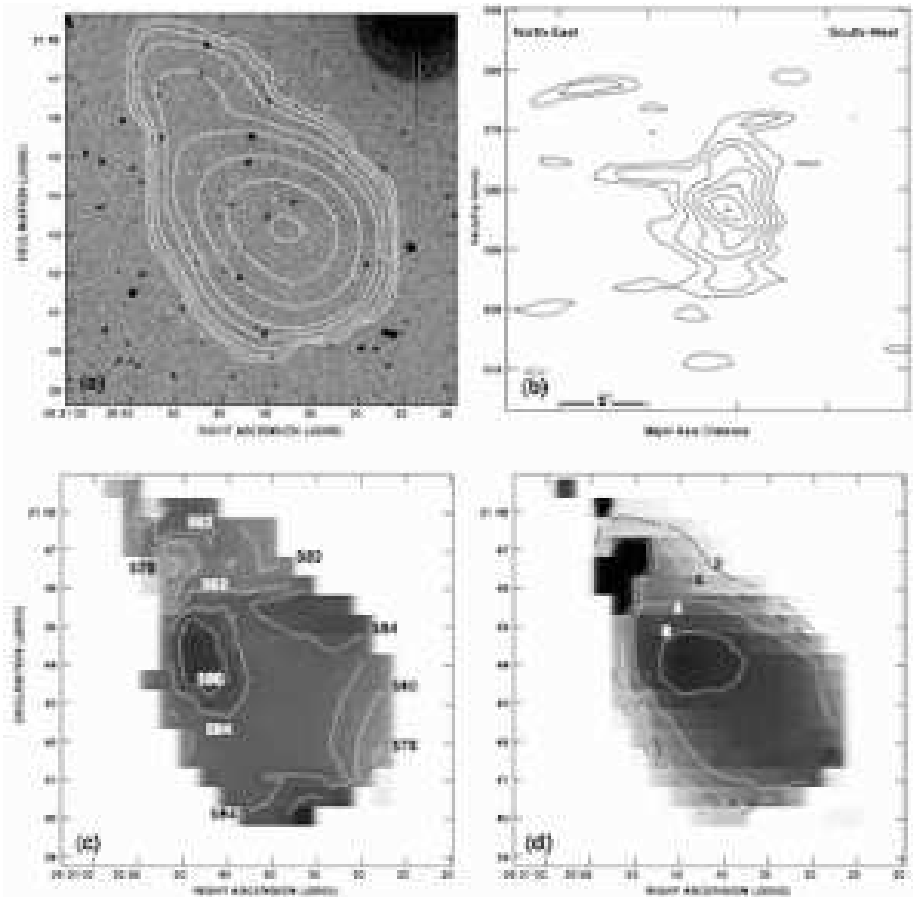}
\caption{H\,I maps of the companion, N2903-HI-1.
(a) Total intensity map, from the RDV-smoothed cubes integrated
over 540 to 617 km s$^{-1}$,
superimposed on the DSS2 Blue image.
Contours are at 2, 5, 10.0, 20, 30, 50, 75, and 100 
Jy beam$^{-1}$ m s$^{-1}$. The column density conversion
is 1.52$\,\times\,$10$^{19}$
cm$^{-2}$ (Jy beam$^{-1}$ km s$^{-1}$)$^{-1}$.  Crosses mark the locations
of all low redshift optical galaxies in the vicinity (Sect.~\ref{2mas})
and the red cross indicates the most probable companion (Table~\ref{table7}).
(b) Position-{v}elocity slice from the RDV-smoothed cubes,
 {averaged over} a 4$^\prime$ wide
swath at a position angle of 45$^\circ$. 
Contours are at  1.0, 1.5, 2.0, 2.5, 3.0, 3.5, and 3.9
mJy beam$^{-1}$.  The angular distance scale is marked at the bottom.
(c) Intensity-weighted mean velocity, from the RD-smoothed cubes, using
the same velocity range as in (a).  
Contours are at 
578, 580, 582, and 584 km s$^{-1}$ and the greyscale ranges from
567 to 593 km s$^{-1}$. (d) Intensity-weighted velocity dispersion, 
from the RD-smoothed cubes, with
contours at 2, 4, 6, and 8 km s$^{-1}$ and using the same 
velocity range as in (a).  The greyscale ranges from 
0.6 to 10 km s$^{-1}$. (The black pixels to the north-east included
noise and are spurious.)
\label{companion_fig}} 
\end{figure*}

Although the true separation and space
velocity of N2903-HI-1 are not known, we can at least adopt the projected
separation and radial velocity offset from NGC~2903 to determine whether the
above speculation is feasible.
Approximating N2903-HI-1 as a sphere,
 its average ISM \HI density,
from the values of Table~\ref{table6}, is
$n_{ISM}\,\approx\,4\,\times\,10^{-4}$ cm$^{-3}$.  
A simple condition for stripping is,
\begin{equation}
\label{ram_pressure_eqn}
n_{halo}\,V_{rel}^2\,>\,n_{ISM}\,G\,{\rm M}_{tot}/R
\end{equation}
where ${\rm M}_{tot}$ is the total mass of N2903-HI-1 and $R$ is
its radius.  Using  ${\rm M}_{tot}\,=\,3\,\times\,10^8$ M$_\odot$,
$V_{rel}\,=\,26$ km s$^{-1}$ (Sect.~\ref{2mas}) and $R$ 
from Table~\ref{table6}, and solving for halo density, we find
$n_{halo} = 2\,\times\,10^{-4}$ cm$^{-3}$.  This value is within the
expected range of halo densities for the Milky Way at the projected
distance of N2903-HI-1 (64 kpc) \citep{grc08}.  
If the relative velocity is larger, stripping will be more effective,
and if the true separation is larger, stripping will be
less effective since the halo density will be lower. 
Nevertheless, this estimate indicates that N2903-HI-1
 could indeed be undergoing
ram pressure stripping under the assumption that the halo of
NGC~2903 resembles the Milky
Way.  The stripping timescale depends on the difference
between the two sides of Eqn.~\ref{ram_pressure_eqn} which is not known.
However, since the magnitudes of the ram pressure and the
 internal energy density
of the companion are similar,
 the stripping timescale may be long, 
consistent with the observation 
of detectable \HI in the 
companion\footnote{If ram pressure {\it did} strongly dominate, then the
 stripping timescale would be minimized, i.e.
$t_{min}\,\approx\,2R/V_{rel}\,=\,3\,\times\,10^8$ yr.}.


It is now interesting to ask how many Local Group dwarf galaxies would
have been detected, if they were distributed around
NGC~2903 similarly to
the Milky Way.  The result is dependent on both their distribution
and \HI content.  
  Seventeen Local Group dwarf galaxies listed in
Mateo (1998)\nocite{mat98} had sufficient data (distances and
\HI data) that we could apply our
$3\,\sigma\,2\,\delta\,V$ detection criterion (Sect.~\ref{noise})
to them.  The result is that
we could have detected 7 of them (41\%) in terms of sensitivity limits.  
However, these 7 galaxies lie at radii between 490 kpc and
1.6 Mpc, in comparison to the effective projected radius of
$R_{eff}$ = 110 kpc (for a circularized field) of our survey.
Since our survey probes a volume that is only 0.7\% of the 
 volume extending to 
1.6 Mpc, it is unlikely that any of these seven galaxies 
would have fallen within our field of view.
Although Mateo does 
not list \HI data for the Large and Small Magellanic Clouds
(LMC and SMC, respectively)
it is clear, however, that we would have detected these systems.
Of the 20 new SDSS dwarf Local Group
galaxies listed in Simon \& Geha (2007)\nocite{sim07}, only one (the 
dwarf irregular, Leo T) has \HI content \citep{rya08}, 
the remainder being dwarf
spheroidals or falling within parameter space intermediate between
dwarf spheroidals and globular clusters. Leo T would
be marginally detectable in our survey but, at a distance of 420
kpc \citep{irw07},
 would also likely lie outside our field of view. 
In spite of our large survey region, therefore, we are still only probing
the inner region of a possible dwarf galaxy population, were it distributed like
our own.

As for HVCs, the HVC population around the Milky Way is not easily
translated to NGC~2903 since HVC distances 
(and therefore \HI masses) are not well known.  Also, \HI column 
densities which are known (typically 10$^{19}$ cm$^{-2}$,
Stanimirovic et al. 2006\nocite{sta06};
Putman et al. 2002\nocite{put02}), will be diluted by large unknown
beam filling factors at the distance of NGC~2903
(234$^{\prime\prime}$ = 10 kpc).
  We can say, however, that the Milky Way's 
HVC Complex C, with a mass of 4.9 $\times$ 10$^6$ M$_\odot$ at a distance
of 10 kpc from the Sun \citep{tho08}, should have been detected,
 provided it were separated in velocity from the bulk of the \HI
in NGC~2903 itself.
 Given the inclination of NGC~2903 and the nature of
HVCs, we expect this criterion to have been met.  Thus, NGC~2903 
lacks such an HVC complex. Indeed, given that our search velocity range 
was over 1000 km s$^{-1}$ (Table~\ref{table4}) and that
 the  median velocity FWHM
of Milky Way HVCs  (36 km s$^{-1}$, 
Putman et al. 2002\nocite{put02}) corresponds to
28 velocity channels in our data, it is surprising that no clear HVC
detections have been made.  Either NGC~2903 lacks HVCs, possibly due
the fact that NGC~2903 is isolated, or its HVCs are of very low mass. 

No dark starless companions have been detected around NGC~2903.
This result is consistent with the ALFALFA survey results
which indicate that all extragalactic
\HI objects can be identified with an optical counterpart
\citep{sai08}.
  The discovery
of one new \HI rich dwarf companion now places the total number of companions
within our surveyed field ($R_{eff}$ = 110 kpc) 
at two: the \HI companion, N2903-HI-1, likely associated with
a dwarf galaxy, 
J093039.96+214324.7, with a dynamical mass 
$\sim$ 3 $\times$ 10$^8$ M$_\odot$ (mean of values in Table~\ref{table6})
and UGC~5806, which is likely a dwarf spheroidal galaxy
(Sect.~\ref{ugc5086_sec}).  Using the SDSS data for the latter galaxy and the same 
transformations as described in Sect.~\ref{2mas}, the stellar mass of UGC~5806
is M$_\star$ = 4.7 $\times$ 10$^7$ M$_\odot$.  Gilmore et al. (2007)\nocite{gil07}
have shown that mass-to-(V-band) light ratios, 
M/$L_V$, for dwarf spheroidal galaxies in
the Local Group range from $\approx$ 4 to 600, with more luminous galaxies
having systematically lower values of M/$L_V$. The absolute magnitude of
UGC~5086 is $M_V$ = -13.7 which is closest to Fornax, amongst Local 
Group dwarfs.  If UGC~5086 has similar properties, then 3 $\le$ 
M/$L_V$ $\le$ 20, implying that 
$1.0\,\times\,10^8$ $\le$ M/M$_\odot$ $\le$ $5.2\,\times\,10^8$.
Adopting the mid-point of this range gives an estimate of 
 $\approx$ 3 $\times$ 10$^8$ M$_\odot$ for the total mass of
UGC~5086.




How many dark matter clumps of total mass  
M$_{tot}\,\gtrsim$ 3$\,\times\,$10$^8$ M$_\odot$
are expected from $\Lambda$CDM predictions?
Since NGC~2903 has a total mass that is similar to
the Milky Way (Sect.~\ref{ngc2903}), we can use the 
Via Lactea simulations of subhalo clumps from 
Diemand et al. (2007a)\nocite{die07b} 
for comparison.  
The fraction of all halo mass that is present in
substructure within a {\it projected} radius of
110 kpc (their Fig. 7) is $f\,=\,0.02$.  With an adopted total halo mass of
1.77$\,\times\,$10$^{12}$ M$_\odot$, this yields a total mass in halo
substructure of M$_{hs}$ = 3.54$\,\times\,$10$^{10}$ M$_\odot$ 
for the projected radius.  
Over the substructure
mass range of 4.6$\,\times\,$10$^{6}$ $\le$ M$_{sub}$/M$_\odot$
 $\le$ 1$\,\times\,$10$^{10}$, the number of clumps per unit mass range is
given by dN/dM$_{sub}\,=\,K/$(M$_{sub})^2$, where $K$ is a constant.  Since
the slope does not change with radius, we can compute $K$ from
M$_{hs}$ = $\int K/({\rm M}_{sub})^2\, {\rm M}_{sub}\, d{\rm M}_{sub}$, finding
 $K\,=\,4.6\,\times\,10^9$ M$_\odot$.  Finally, we compute the expected
number of clumps with masses greater than M$_{sub}$ from
N($>$ M$_{sub}$)\,=\,K/M$_{sub}$.  For M$_{sub}$ = 
$3\,\times\,$10$^8$ M$_\odot$, this yields 15 clumps in comparison to the
two observed.  
If the dynamical mass of N2903-HI-1 is higher than 
$3\,\times\,$10$^8$ M$_\odot$ (for example, if it is rotating and
nearly face-on), then the discrepancy reduces.  However, its mass 
would have to be as high as $\sim\,5\,\times\,10^9$ M$_\odot$ for there to
be agreement with the theoretical expectation.  Such a high 
value would require
N2903-HI-1 to represent a rotating system
in a special geometry
with an inclination less than 10$^{\circ}$
(Sect.~\ref{2mas}). 
 Again, although this possibility
cannot be ruled out completely, it is quite unlikely.
Thus, it would appear that there is a discrepancy between
the expected number of companions and the number observed.  

\begin{figure}
\plotone{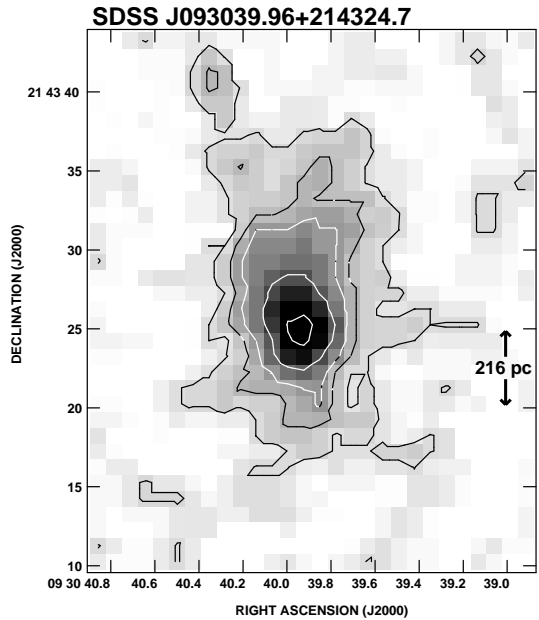} 
\caption{The galaxy, SDSS J093039.96+214324.7, which is the
most likely optical counterpart to NGC~2903-HI-1.   The image is
from the DSS2 Blue image and contours are shown in arbitrary units, with
the first contour set at the 2$\,\sigma$ level.  The scale,
for a distance equal to that of NGC~2903, is shown at right. 
\label{dss2_optical_companion}}
\end{figure}

Since we have a detection threshold that is lower
still than the value of the detected companion, we can also ask how
many dark matter clumps would we expect to see, 
were they \HI rich.  For example,
our lowest \HI detection threshold is 2$\,\times\,$10$^5$ M$_\odot$
(Table~\ref{table4}).  If M$_{tot}$/M$_{HI}$ $\approx$ 100 for
a dwarf galaxy in the field (equivalent to the N2903-HI-1 value), then 
we could have detected companions of total mass as low as 
2$\,\times\,$10$^7$ M$_\odot$
via their \HI.  Therefore,
within our field of view, using the above relation,
 we should have detected 230 galaxies as opposed to the one observed.
 This is strongly discordant with $\Lambda$CDM.  The
conclusion is that, if this many dark matter clumps exist in the region, then
they are clearly not \HI rich, i.e. they contain no \HI or they contain
\HI at a level lower
than 1\%.

\section{Conclusions}
\label{conclusions}

Using the Arecibo telescope with the ALFA receiver, we have mapped
NGC~2903 and its environment with very high sensitivity and 
coverage.
 Our lowest point source detection limit 
is $2\,\times\,10^5$ M$_\odot$ and almost 
40 thousand
square kpc of sky has been fully covered.
The Arecibo ALFA beams have been carefully characterized as a function
of azimuth,
allowing us
to clean 
each beam as a function of azimuth from the
\HI data cube.  
With a velocity coverage of 1035 km s$^{-1}$ and fine velocity
resolution (2.6 km s$^{-1}$), our combination of observing parameters
makes this survey unique and among the most sensitive
and complete of a nearby galaxy.  Although details of NGC~2903, itself,
are left to future work, our results
show that the \HI envelope around NGC~2903 is much larger
than previously known, extending to at least 3 times the optical galaxy
diameter.

The fact that we have targeted an apparently isolated, non-interacting
galaxy to a very low sensitivity limit has clearly been an advantage in 
 the search for \HI companions.  The discovery of only one isolated
\HI companion, N2903-HI-1,
which appears to have a small optical counterpart,
 is a significant result.  The optical companion is likely a dwarf
galaxy with a stellar mass approximately equal to its \HI
mass with the \HI in a broad envelope, approximately 
8 times larger, around it.  
The best estimate of its dynamical mass is $3\,\times\,10^8$ M$_\odot$. 
We have no convincing HVC detections.

In the field surveyed, there are now two known companion galaxies, our
new discovery as well as what is likely a dwarf spheroidal galaxy, UGC~5086, 
the latter with a total mass likely
comparable to N2903-HI-1.  In this region, $\Lambda$CDM scenarios 
(specifically, the Via Lactea model)
predict
15 companions for a Milky Way-type galaxy,
with masses greater than $3\,\times\,10^8$ M$_\odot$.
Given our \HI detection limits, however, {\it if} companions to NGC~2903
contained \HI at the 1\% level in comparison to their total masses, then
we should have detected 230 of them.
If these clumps are present as predicted, they do not contain appreciable
\HI.  They may be starless dark clumps or very low luminosity dark-matter
dominated dwarf spheroidals.




\acknowledgments
We are grateful to students K. Marble of Queen's University and
I. Bah, A. Altaf, and J. Goldstein of Lafayette College for their
assistance with the data reductions. 
Many thanks also to 
P. Perillat and the Arecibo staff for their knowledge and assistance. 
JAI gratefully acknowledges a grant from the Natural Sciences and Engineering
Research Council of Canada.  GLH gratefully acknowledges grants from the
Lafayette College Academic Research Committee.
This research has made use of the NASA/IPAC Extragalactic Database (NED) 
which is operated by the Jet Propulsion Laboratory, California Institute 
of Technology, under contract with the National Aeronautics and Space 
Administration.
Funding for the SDSS and SDSS-II has been provided by the Alfred P. 
Sloan Foundation, the Participating Institutions, the National 
Science Foundation, the U.S. Department of Energy, the National 
Aeronautics and Space Administration, the Japanese Monbukagakusho, 
the Max Planck Society, and the Higher Education Funding Council 
for England. The SDSS Web Site is http://www.sdss.org/.



{\it Facilities:} \facility{Arecibo}.

\clearpage

\end{document}